\documentclass[usenatbib]{mnras}

\usepackage{graphicx}% Include figure files
\usepackage{dcolumn}% Align table columns on decimal point
\usepackage{amssymb} %maths
\usepackage{amsmath}
\usepackage{ctable}
\usepackage{bm}% bold math
\usepackage{epsfig}
\usepackage{epstopdf}
\usepackage{epsf,color}
\usepackage{comment}
\usepackage{aas_macros}
\usepackage{url}
\usepackage{widetext}
\usepackage{color, colortbl}
\newcommand{\cmnt}[1]{}

\def\hatn{\mathbf{\hat n}}

\title{Calibrating magnification bias for the $E_G$ statistic to test general relativity}

\author[S.~Yang et al.]{Shengqi Yang\thanks{Email: sy1823@nyu.edu}$^1$ and Anthony R. Pullen$^1$\\
$^1$Center for Cosmology and Particle Physics, Department of Physics, New York University, 726 Broadway, New York, NY, 10003, U.S.A.}

\date{Accepted XXX. Received YYY; in original form ZZZ}

\pubyear{2016}

\begin{document}
\label{firstpage}
\pagerange{\pageref{firstpage}--\pageref{lastpage}}
\maketitle

\begin{abstract}
We assess the effect of magnification bias on the $E_G$ statistic for probing gravity. $E_G$, a statistic constructed from power spectrum estimates of both weak lensing and redshift space distortions (RSD), directly tests general relativity (GR) while in principle being independent of clustering bias.  This property has motivated its recent use in multiple tests of GR. Recent work has suggested that the magnification bias of galaxies due to foreground matter perturbations breaks the bias-independence of the $E_G$ statistic. The magnitude of this effect is very sensitive to the clustering and magnification biases of the galaxy sample. We show that for realistic values of the clustering and magnification biases, the effect for magnification bias is small relative to statistical errors for most spectroscopic galaxy surveys but large for photometric galaxy surveys. For the cases with significant magnification bias, we propose a method to calibrate magnification bias using measurements of the lensing auto-power spectrum. We test this calibration method using simulations, finding that our calibration method can calibrate $E_G$ from 2-4 times the simulation error to well within the errors, although its accuracy is sensitive to the precision of the measured redshift distribution and magnification bias, but not the clustering bias. This work gives strong evidence that this method will work increasingly well in future CMB lensing surveys.
\end{abstract}
\begin{keywords}
cosmology: theory, cosmology: observations -- gravitation -- gravitational lensing: weak -- large scale structure of the universe -- cosmic microwave background
\end{keywords}
%\maketitle

\section{Introduction}

The discovery of cosmic acceleration \citep{Riess1998,Perlmutter1999} is a phenomenon in search of a theoretical explanation. On one hand, the acceleration can be explained by introducing dark energy which provides negative pressure \citep{Peebles2002}. On the other hand, on cosmological scales, the dynamics of gravity are untested and may deviate from general relativity (GR). Various models with modified gravity \citep{Dvali2000,Carroll2004} are degenerate with dark energy models and can also explain the cosmic acceleration. One of the major interests of large-scale structure (LSS) surveys is to break the degeneracy between these two explanations. \par
An additional gravitational process that can break this degeneracy is the growth of structure. The growth rate, measured through redshift space distortions (RSD) \citep{Kaiser1987b,Hamilton1997}, has been constrained by various analyses \citep{Beutler2012,Samushia2013,Alam2015,2018arXiv180102891R,2018MNRAS.477.1639Z,2018arXiv180102656H,2018MNRAS.477.1604G,2018arXiv180103043Z}.  However, the growth rate estimated from its effect on the power spectrum suffers from a degeneracy with clustering bias and the scalar amplitude $A_s$, or equivalently $\sigma_8$.  Since the clustering bias is very galaxy sample-specific and not simple to measure, an estimate of the growth rate independent of clustering bias is advantageous.  \par
To remedy this, \citet{Zhang2007} introduce the $E_G$ statistic to probe gravity. $E_G$ is a statistic constructed from the ratio between Laplacian of the sum of two curvature fields $\nabla^2(\Phi+\Psi)$ and the velocity divergence field $\theta=\nabla\cdot\vec{v}/H(z)$, where $\vec{v}$ is the comoving peculiar velocity. While $\nabla^2(\Phi+\Psi)$ is probed through lensing measurements, originally through galaxy lensing, $\theta$ is probed through the RSD measurements of the growth rate.  Not only is this statistic independent of clustering bias on large scales, but also GR predicts this ratio to be scale-independent.  If $E_G$ is measured to be scale-dependent, and not in agreement with the prediction from $\Lambda$CDM cosmogony, then modified gravity would become a promising solution for cosmic acceleration.  In addition, various factors which may contribute to uncertainty of $E_G$ were analyzed in \cite{Leonard2015}. $E_G$ was first measured by \citet{Reyes2010a} at redshift $z=0.32$ using galaxy-galaxy lensing information traced by the Luminous Red Galaxy (LRG) sample \citep{Eisenstein2001} from the Sloan Digital Sky Survey (SDSS) \citep{York2000}. \cite{Blake2015} then measured $E_G$ at redshifts $z=0.32$ and $z=0.57$ using galaxy lensing information from multiple datasets. \cite{Giannantonio2016} measure $D_g$, a modified $E_G$ in which the growth rate is not used in the estimator. \cite{Pullen2015} discuss that $\nabla^2(\Phi+\Psi)$ can be estimated through CMB lensing cross-correlated with galaxies to allow measurement of $E_G$ at higher redshifts, a measurement which was first demonstrated in \cite{Pullen2016}.\par 
All previous measurements of $E_G$ have assumed that the galaxy sample is only correlated with the lensing perturbations \emph{at that redshift}.  However, \cite{MoradinezhadDizgah2016a}, hereafter D16, point out that the magnification bias, in which foreground density perturbations both magnify the regions around galaxies and distort the selection of galaxies near the survey flux limit, can bias the galaxy-convergence and galaxy-galaxy angular power spectra. It was predicted in D16 that $E_G$ could deviate by around 2\% for current surveys but up to $40\%$ from the theoretical value for upcoming high-redshift surveys.  Recently, \cite{Singh2018} consider magnification bias for an $E_G$ measurement using galaxy lensing and CMB lensing in real space with data from CMASS and LOWZ \citep{2015ApJS..219...12A}, also finding an expected deviation $\sim2$\%.

It is worth mentioning that since surface brightness is conserved, an $E_G$ measurement is free from magnification bias if the galaxy survey is replaced by an intensity mapping survey in the $E_G$ estimator \cite{2016MNRAS.461.1457P}. However, measuring $E_G$ through 21 cm emission is unlikely to be possible because continuum foregrounds and calibration errors may prevent the measurement of the global intensity signal to the accuracy necessary to measure the growth rate. It may still be possible to measure $E_G$ through CO rotation line and CII line since they are less contaminated by foreground.\par

In this paper, we test how significantly magnification bias will affect upcoming measurements of $E_G$, specifically $E_G$ estimated from CMB lensing.
%However, we find this work does not apply extra redshift and bias calibration, which is not proper for the broad redshift distribution it uses. Also, D16 does not apply a realistic galaxy bias nor realistic redshift distribution ranges, which will cause a discrepancy between its prediction and real measurement. More importantly, this work uses inconsistent sign conventions for two scalar fields, computing a negative galaxy-convergence cross angular power spectrum. One aim of this paper is to show that, after correcting the sign, applying realistic bias values, redshift distribution ranges and repeating the work of D16, the $E_G$ deviation due to lensing magnification effect is not as large as previously predicted.
We start by giving the derivation for the magnification bias effect on $E_G$, finding that the bias is very sensitive to the assumed clustering and magnification biases.  We then apply redshift and scale-dependent clustering bias calibrations proposed by \cite{Pullen2016} based on calibrations from \citet{Reyes2010a} to study how $E_G$ is biased for CMB lensing maps from Planck \citep{Lawrence2015} and Advanced ACT (Adv.~ACT) \citep{Henderson2015} cross-correlated with both spectroscopic and photometric galaxy surveys, including the Baryon Oscillation Spectroscopic Survey (BOSS) \citep{2013AJ....145...10D} and the Extended Baryon Oscillation Spectroscopic Survey (eBOSS) \citep{2016AJ....151...44D}, the Dark Energy Spectroscopic Instrument (DESI) \citep{Levi2013}, the Dark Energy Survey (DES) \citep{2005astro.ph.10346T}, the Large Synoptic Survey Telescope (LSST) \citep{2012arXiv1211.0310L}, and Euclid \citep{Laureijs2011}.  We confirm that the proposed CMB ``Stage 4'' survey (CMB-S4) \citep{Abazajian2016} does not significantly decrease the errors on $E_G$ further, and thus do not display specific results from this survey. We find through theoretical analysis and simulations that the deviation of $E_G$ should be less than 5\% in quasi-linear scales for spectroscopic and photometric galaxy surveys.  We find that most spectroscopic galaxy surveys, except for the DESI luminous red galaxy (LRG) and emission line galaxy (ELG) surveys cross-correlated with CMB lensing maps from Adv.~ACT, do not have magnification biases large enough to significantly contaminate $E_G$.  On the other hand, we find that the photometric galaxy surveys, including DES, LSST, and Euclid, will have large magnification biases that will contaminate $E_G$.  In these cases, we propose a method to clean the higher-order terms contributed by the lensing magnification effect with observable angular power spectrum, finding it calibrates $E_G$ effectively. We believe this calibration will work well in future surveys with higher precision.\par
The plan of this paper is as follows: In section \ref{S:form} we briefly review the $E_G$ estimator, present our formalism for the magnification bias of $E_G$, and construct our calibrations to remove the bias. Section \ref{S:results} uses simulations to test its performance of our calibration method as well as its stability relative to errors in modeling the calibration. We conclude in section \ref{S:conclu}. In this work we assume the following cosmological parameters: $\Omega_m=0.309$, $\Omega_bh^2=0.02226$, $\Omega_ch^2=0.1193$, $\Omega_k=0$, $h=0.679$, $n_s=0.9667$, and $\sigma_8=0.8$ given in \cite{PlanckCollaboration2015}, \cite{Alam2016}.

\section{$E_G$ Formalism and Calibration Method}\label{S:form}

The first part of this section derives the formalism for the distortion of $E_G$ due to magnification bias.  Although this is mostly derived in D16, we do make some small corrections to their derivation that do not on their own affect the final results for $\Delta E_G/E_G$. In the second part of this section we present our method to calibrate the magnification bias in $E_G$, which we test in Section \ref{S:results}.

\subsection{$E_G$ statistic review} \label{S:review}\par

The expression for $E_G$ without magnification bias was derived previously in the literature, so we just review the main points here; see \citet{Pullen2015} and \citet{Pullen2016} for a complete review.  Note that we assume flat space for all $E_G$ analyses.  $E_G$ is defined in Fourier space for any metric theory of gravity \citet{Zhang2007} as
\begin{eqnarray}
    E_G\equiv\frac{c^2k^2(\Phi+\Psi)}{3H_0^2(1+z)\theta(k)}\, ,
\end{eqnarray}
where $H_0$ is the Hubble expansion rate today, $\Phi$ and $\Psi$ are the weak-field potentials in the space and time metric components, respectively, and $\theta$ is the velocity divergence perturbation. Both the numerator and denominator can be related to the matter overdensity $\delta_m=\delta\rho/\bar{\rho}$: the potentials through the field equations of the theory and $\theta=f(z)\delta_m$, where $f(z)$ is the growth rate.  For GR, we write
\begin{eqnarray}
    E_G=\frac{\Omega_{m0}}{f(z)}\, ,
\end{eqnarray}
where $\Omega_{m0}$ is the relative matter density today, while for other theories of gravity the expression for $E_G$ will be altered and potentially even attain scale-dependence on large scales.

We state that $E_G$ can be estimated \citep{Pullen2015} as 
\begin{eqnarray}
    \hat{E}_G(\ell,\bar{z})=\Gamma(\bar{z})\frac{C_\ell^{\kappa g}}{\beta(\bar{z}) C_\ell^{gg}}\, ,
\end{eqnarray}
where $\bar{z}$ is the mean redshift of the galaxy survey, $C_\ell^{\kappa g}$ is the measured lensing convergence-galaxy angular cross-power spectrum, $C_\ell^{gg}$ is the measured galaxy angular auto-power spectrum,  $\beta=f(z)/b(z)$, where $b(z)$ is the clustering bias as a function of redshift $z$ and $\beta$ is usually fitted from measurements of the monopole and quadrupole moments of the anisotropic correlation function $\xi(r,\mu)$, and $\Gamma(\bar{z})$ is a calibration factor.  The expressions for $C_\ell^{\kappa g}$ and $C_\ell^{gg}$ can be written as
\begin{eqnarray}\label{E:clab}
    C_\ell^{AB}=\frac{2}{\pi}\int_0^{\infty}k^2dkP(k,z=0)W_\ell^A(k)W_\ell^B(k)\, ,
\end{eqnarray}
where $P(k,z)$ is the matter power spectrum and $W_\ell^A$ is a window function for observable $A$. The window functions for both lensing convergence $\kappa$ and galaxy overdensity $\delta_g$ (assuming no magnification bias) are given by
\begin{eqnarray}
        W_\ell^{\kappa}(k,z)&=&\frac{3\Omega_{m0}H_0^2}{2c^2}\int_0^zdz'\frac{c}{H(z')}W(z,z')D(z')\nonumber\\
        &&\times(1+z')j_\ell(k\chi(z'))\, ,
\end{eqnarray}
and
\begin{eqnarray}
    W_\ell^g(k)=\int dz\,f_g(z)b(z)D(z)j_\ell(k\chi(z))\, ,
\end{eqnarray}
where $\chi$ is the comoving distance given by
\begin{equation}
\chi=\int_0^z\dfrac{cdz'}{H(z')}\, ,
\end{equation}
$f_g$ is the galaxy redshift distribution, $D(z)$ is the normalized growth factor, and $W(z,z')$ is the lensing kernel for a lens redshift $z'$ and a source redshift $z$ written as
\begin{eqnarray}
    W(z,z')=\chi(z')\frac{\chi(z)-\chi(z')}{\chi(z)}\, .
\end{eqnarray}
Note that unlike most treatments we keep the source redshift explicit, which will be helpful later.  The calibration factor $\Gamma$ is given by
\begin{eqnarray}
\Gamma(z)=C_\Gamma C_b\frac{2c}{3H_0}\left[\frac{H(z)f_g(z)}{H_0(1+z)W(z_{CMB},z)}\right]\, , 
\end{eqnarray}
where $C_\Gamma$ and $C_b$ are extra calibrations for the broad redshift distribution and lensing kernel and for the scale-dependent bias due to nonlinear clustering, respectively, given by
\begin{eqnarray}\label{eq:22}
        C_{\Gamma}(\ell,z)&=&\frac{W(z_{CMB},z)(1+z)}{2f_g(z)}\frac{c}{H(z)}\frac{C^{mg}_\ell}{Q^{mg}_\ell}\nonumber\\
        C_b(\ell)&=&\frac{C_\ell^{gg}}{b(\bar{z})C_\ell^{mg}}\, ,
\end{eqnarray}
where
\begin{eqnarray}
        C_\ell^{mg}&=&\int_{0}^{\infty}dz\frac{H(z)}{c}f_g^2(z)\chi^{-2}(z)P_{mg}\left(k=\frac{\ell}{\chi(z)},z\right)\nonumber\\
        Q^{mg}_\ell&=&\frac{1}{2}\int_{0}^{\infty}dzW(z_{CMB},z)f_g(z)\chi^{-2}(z)(1+z)\nonumber\\
        &&\times P_{mg}\left(k=\frac{\ell}{\chi(z)},z\right)\, ,
\end{eqnarray}
and $P_{mg}(k,z)=b(z)P(k,z)$.

\subsection{$E_G$ distortion due to magnification bias} \label{S:bias}\par

The effect of magnification bias on $C_\ell^{\kappa g}$ and $C_\ell^{gg}$ are given in D16, so we do not give a full derivation here.  Instead, we just state the main points.  The correct expression for the observed galaxy overdensity $\Delta_g$ in terms of the matter overdensity $\delta_m$, including only magnification bias and not other relativistic effects, is given by
\begin{eqnarray}
    \Delta_g(\hatn,z)=b(z)\delta_m(\hatn,z)+(5s-2)\kappa(\hatn,z)\, ,
\end{eqnarray}
where $\kappa(\hatn,z)$ is the lensing convergence map with a source redshift $z$ and $s$ is the slope of the cumulative magnitude function, also called the \emph{magnification bias}, given by
\begin{eqnarray}\label{eq:13}
    s=\left.\frac{d\log_{10} n_g(m<m_*)}{dm}\right|_{m=m_*}
\end{eqnarray}
where $n_g(m<m_*)$ is the number density of galaxies with an apparent magnitude $m$ less than $m_*$.  Note that the magnification bias term differs from the same in Eq. 2.13 in D16 by a minus sign, although it does not affect the magnitude of $\Delta E_G/E_G$ because the minus cancels with a second minus sign error in D16 for $C_\ell^{\kappa g}$.  Also, we see here the well-known result that the magnification bias vanishes for $s=0.4$ which will be important in our analysis.  Substituting $\delta_g$ for $\Delta_g$, we find that the window function for $W_\ell^g$ gains an extra term such that $W_\ell^g=W_\ell^{g1}+W_\ell^{g2}$ where $W_\ell^{g1}$ is the expression without magnification bias and
\begin{eqnarray}
    W_\ell^{g2}(k)=(5s-2)\int dz\,f_g(z)W_\ell^\kappa(k,z)\, .
\end{eqnarray}
Putting them back to the general expression of angular power spectrum gives
\begin{eqnarray}\label{eq:17}
        C_\ell^{gg}=C_\ell^{g1g1}+2C_\ell^{g1g2}+C_\ell^{g2g2}
\end{eqnarray}
where
\begin{eqnarray}\label{E:gg}
    C_\ell^{gigj}=\frac{2}{\pi}\int_0^{\infty}k^2dkP(k,z=0)W_\ell^{gi}(k)W_\ell^{gj}(k)\, ,
\end{eqnarray}
and
\begin{eqnarray}\label{eq:18}
        C_\ell^{\kappa g}&=&C_\ell^{\kappa g1}+C_\ell^{\kappa g2}
\end{eqnarray}
where
\begin{eqnarray}\label{E:kg}
        C_\ell^{\kappa gi}&=&\frac{2}{\pi}\int_0^{\infty}k^2dkP(k,z=0)W_\ell^{gi}W_\ell^{\kappa}(k,z*)
\end{eqnarray}
Note that to compute these $C_\ell$s we will use the Limber approximation \citep{Loverde2008a}, which works well on scales $l+1/2\gtrsim\bar{r}/\sigma$, where $\bar{r}$ and $\sigma$ are the peak and width of radial selection function. Since $\bar{r}/\sigma$ for all the redshift distributions and lensing kernels we consider in this work is less than 10, to avoid expensive spherical Bessel function integration we apply the Limber approximation to compute angular power spectra at scales $l>10$. Note that scales $l<10$ will not be used in this work.  We present the Limber approximation expressions in Appendix \ref{S:cllimber}.

These expressions now allow us to predict how significantly magnification bias will distort our measurement.  While we will consider this for specific surveys in the next section, here we consider a more general case.  D16 performed a similar analysis, and found that $C^{gg}_{12}$, $C^{gg}_{22}$ and $C^{\kappa g}_2$ contributed by magnification bias could considerably distort $E_G$ and ruin its independence from clustering bias.\par
We consider redshift distributions identical to those in D16:
\begin{eqnarray}\label{eq:20}
    f(z)\propto\left(\frac{z}{z_g}\right)^2\exp\left[-\left(\frac{z}{z_g}\right)^2\right], z_g=0.57 \nonumber\\
    f(z)=\frac{1}{\sigma\sqrt{2\pi}}\exp\left(-\frac{(z-z_g)^2}{2\sigma^2}\right), z_g=1.5, \sigma=0.68\, .
\end{eqnarray}
 where we limit the redshift range to $z\in[0.47-0.67]$ for the first distribution and $z\in[1.4,1.6]$ for the second one as analogs of BOSS+eBOSS LRG and DESI ELG survey, respectively. In both cases, we assume a clustering bias $b=1$ and $s=0$ for simplicity as in D16. %In this section we use a realistic galaxy bias $b=2.3$ instead. 
 We use the notation:
\begin{equation}\label{eq:19}
\begin{split}
    &E_{G0}=\Gamma(\bar{z})\frac{C^{\kappa g}_1}{\beta(\bar{z})C^{gg}_{11}}, E_{G1}=\Gamma(\bar{z})\frac{C^{\kappa g}}{\beta(\bar{z})C^{gg}}\\
    &\Delta C^{\kappa g}_l=C^{\kappa g}_{l1}-C^{\kappa g}_l\\
    &\Delta C^{gg}_l=C^{gg}_{l11}-C^{gg}_{l}\\
    &\Delta E_G=E_{G0}-E_{G1}
\end{split}
\end{equation}\par
\begin{figure*}
	\centering
		\includegraphics[width = 7cm, height=6cm]{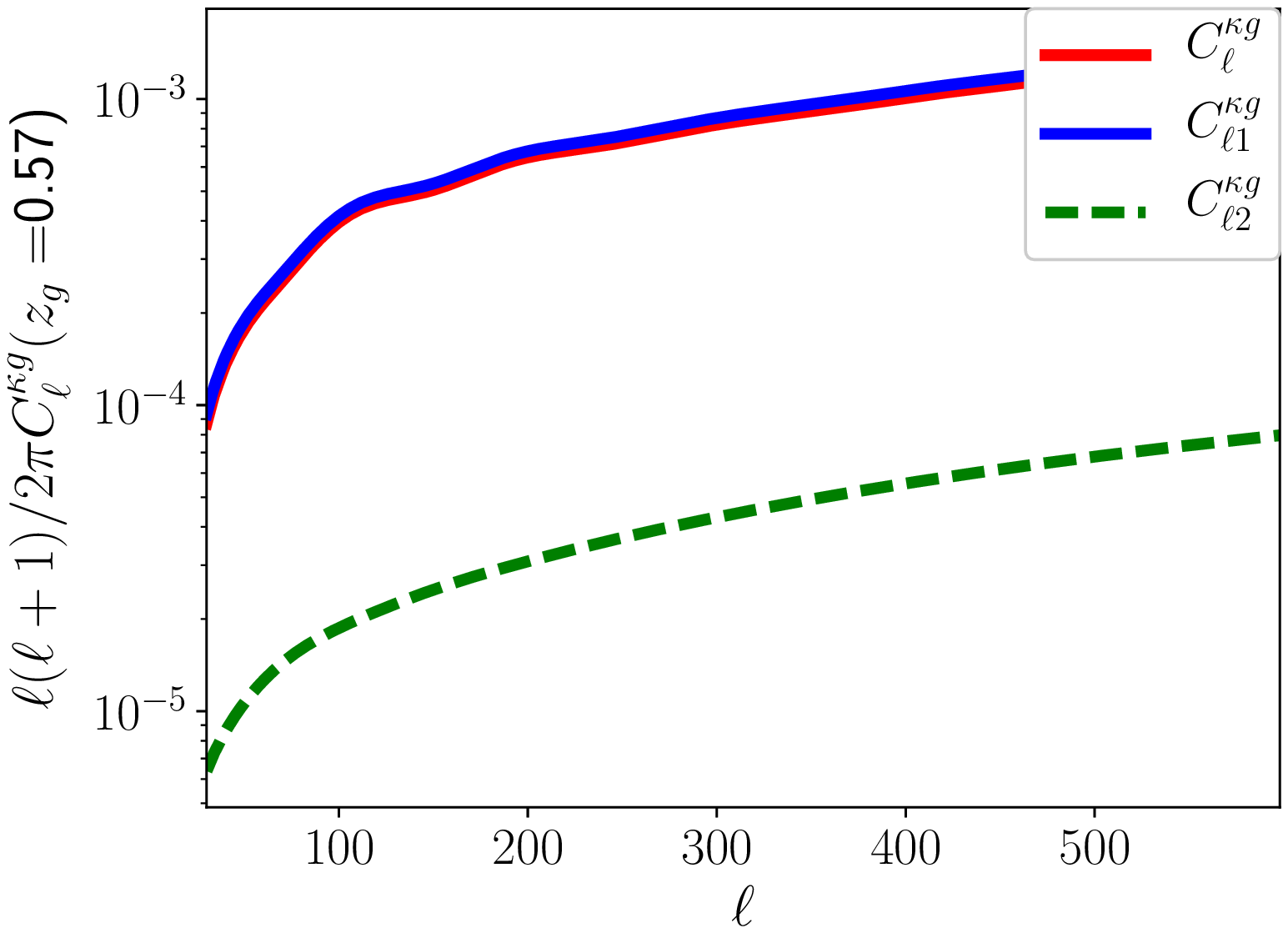}
		\includegraphics[width = 7cm, height=6cm]{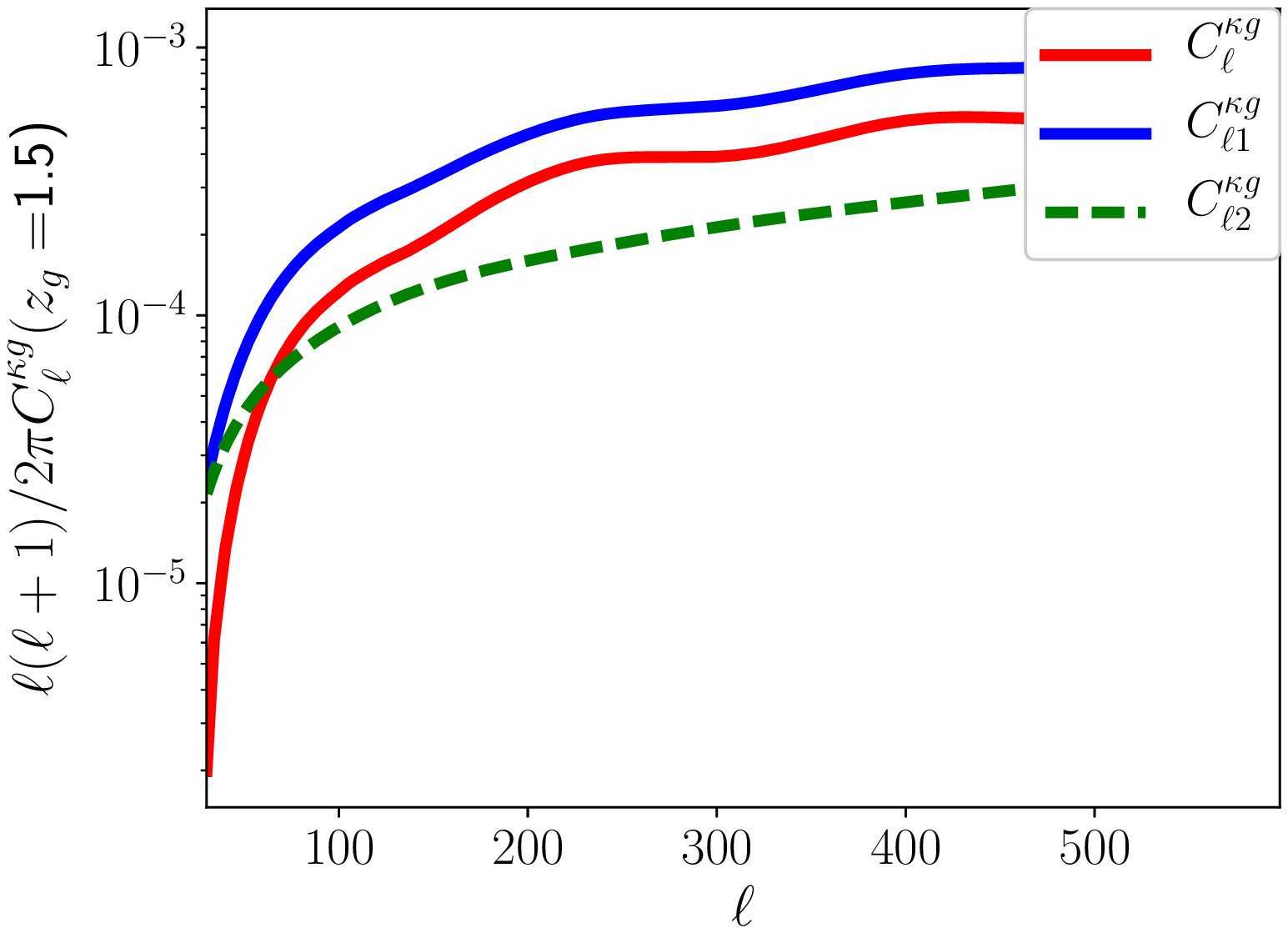}\\
		\includegraphics[width = 7cm, height=6cm]{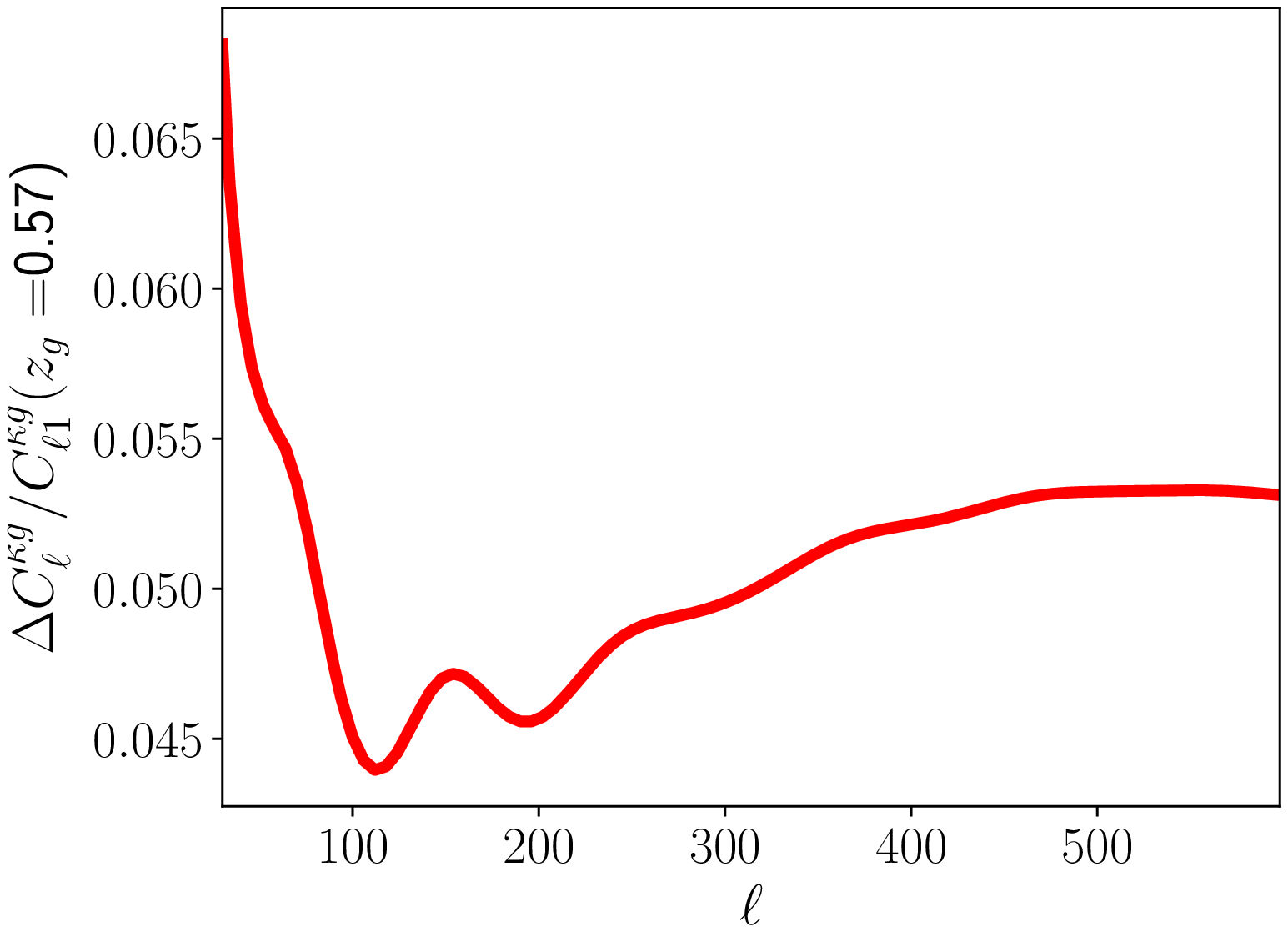}
		\includegraphics[width = 7cm, height=6cm]{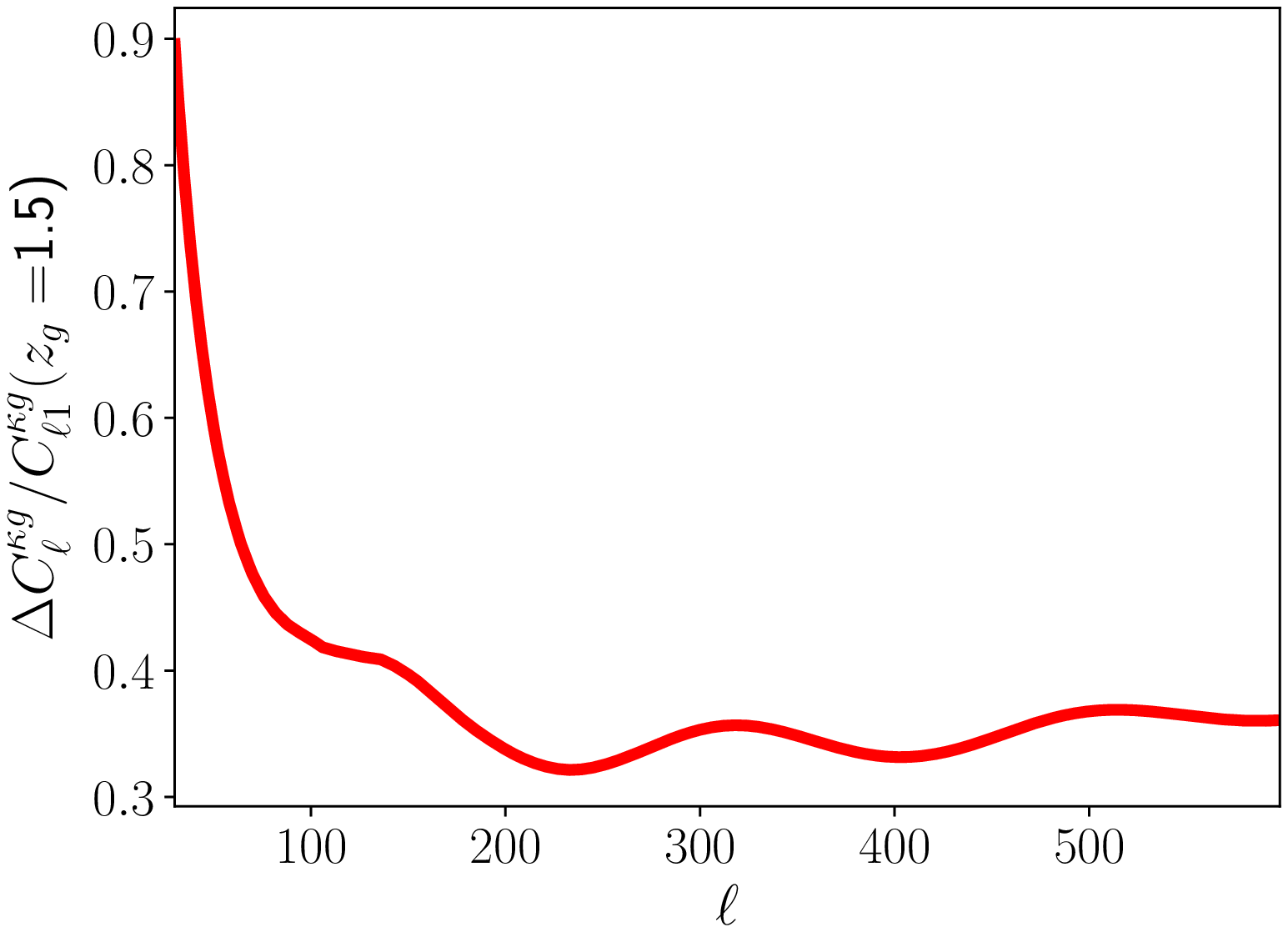}\\
	\caption{$C^{\kappa g}_l$ deviation at redshifts centered at $z=0.57$ (left) and $z=1.5$ (right). The first row shows the absolute deviation and the second row shows the fractional deviation. Dashed line denotes negative correlations. Galaxy bias and magnification bias parameter are set as $b=1$ and $s=0$ in the calculation.}
	\label{fig:1}
\end{figure*}

  The deviations for $C^{gg}_\ell$ and $C^{\kappa g}_\ell$ are shown in Figure \ref{fig:1}.  While these results are not exactly the same as in D16, they are broadly similar. One source of the small deviations between our results and those from D16 is the different cosmological parameters we use. Also, D16 does not specify the redshift distribution range it uses. In this section we integrate for angular power spectrum over redshift range $z\in[\bar{z}-0.1,\bar{z}+0.1]$, which might be different to what D16 uses and could cause disagreement between our results. As mentioned in D16, we expect choosing different $b$ and $s$ values will change the deviations in $C_\ell^{\kappa g}$ and $C_\ell^{gg}$ by the factor $(2-5s)/(2b)$, which we will consider further in regards to the fractional deviation of $E_G$.\par%  We see that for more narrow redshift distributions, the magnification bias effects are much smaller.  This result is because the bias on $W_\ell^g$ is dependent on the redshift range of the target galaxies.  In particular the CMB lensing is most efficient at $z\simeq 2$, meaning redshifts higher than our constricted ranges will contribute significantly to the bias \Ant{Guessing, need to check: The result shows redshift range influence $\Delta E_G/E_G$ in a non-trivial way.}.  In addition, we expect choosing different $b$ and $s$ values will change the deviations in $C_\ell^{\kappa g}$ and $C_\ell^{gg}$ by the factor $(2-5s)/(2b)$.  For a more reasonable $b=2.3$ and $s=XX$, this will change the deviations by a factor of XXX.\Ant{Setting s=0, b=1: $\langle\Delta E_G/E_G(\bar{z}=0.57)\rangle=4.55\%$, $\langle\Delta E_G/E_G(\bar{z}=1.5)\rangle=42.4\%$. Setting s=0.2, b=2.3: $\langle\Delta E_G/E_G(\bar{z}=0.57)\rangle=0.94\%$, $\langle\Delta E_G/E_G(\bar{z}=1.5)\rangle=8.88\%$. Setting s=0.48, b=2.3: $\langle\Delta E_G/E_G(\bar{z}=0.57)\rangle=-0.37\%$, $\langle\Delta E_G/E_G(\bar{z}=1.5)\rangle=-3.47\%$}\par
\par 

\begin{figure*}
	\centering
		\includegraphics[width = 7cm, height=6cm]{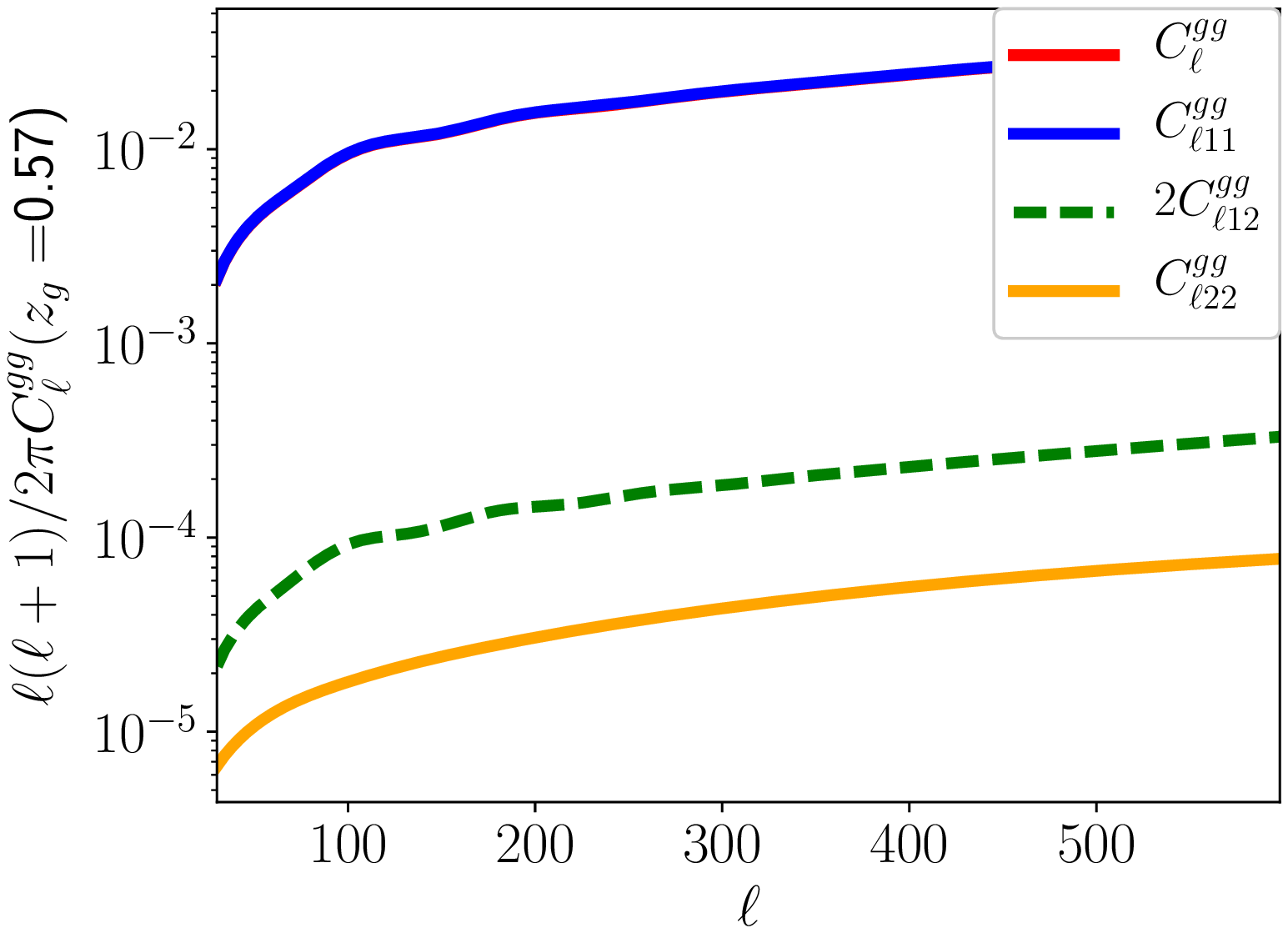}
		\includegraphics[width = 7cm, height=6cm]{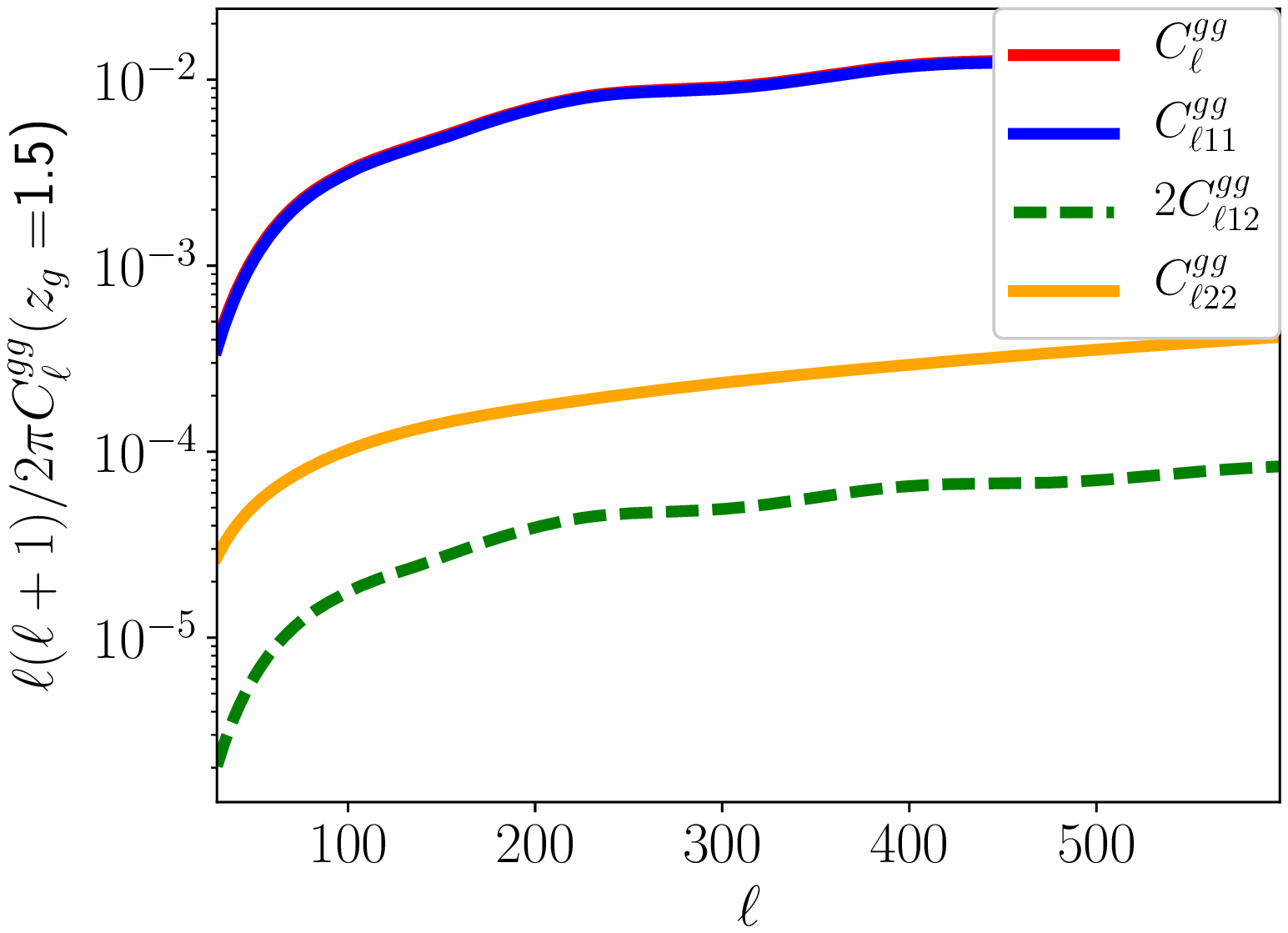}\\
		\includegraphics[width = 7cm, height=6cm]{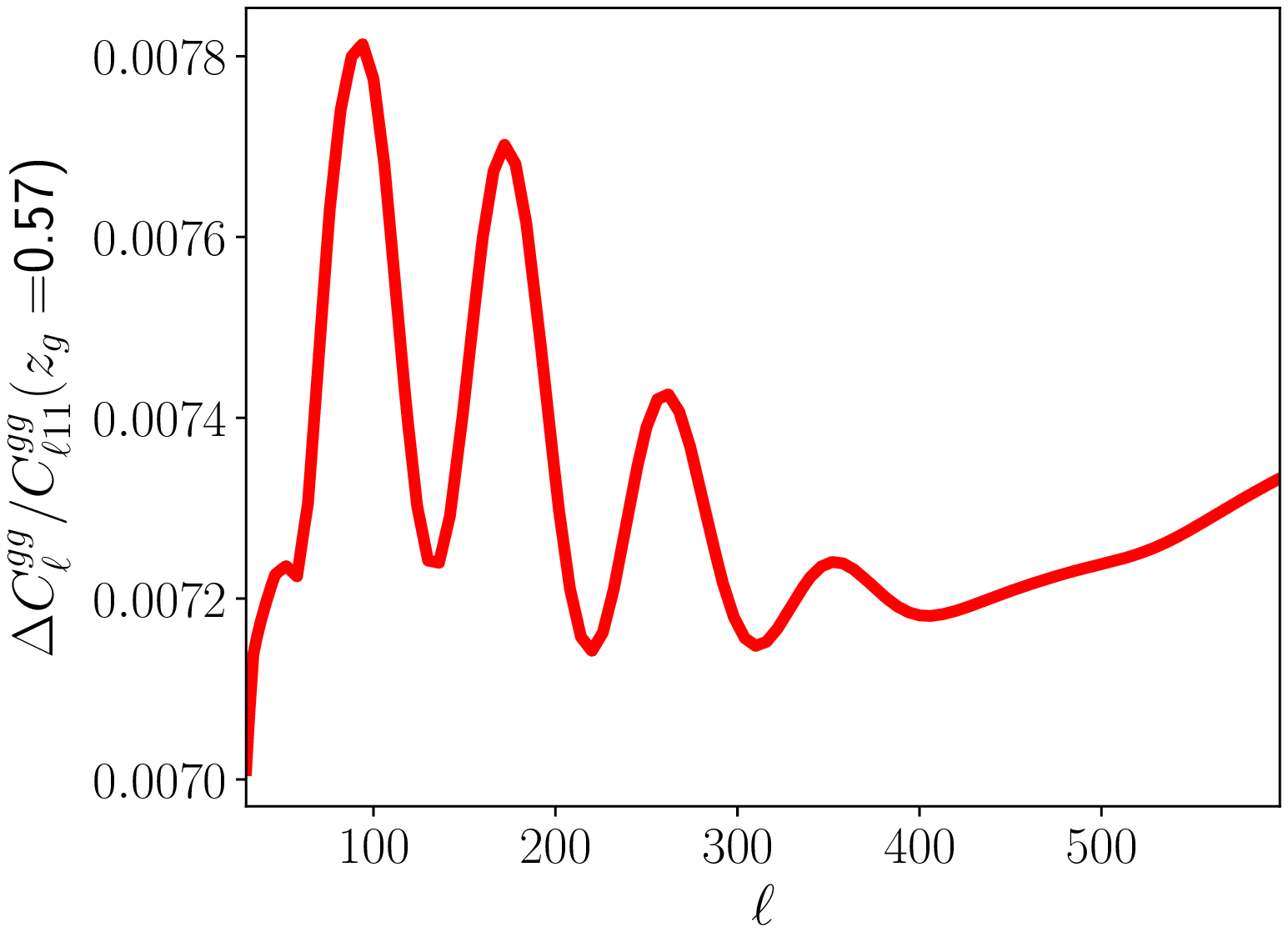}
		\includegraphics[width = 7cm, height=6cm]{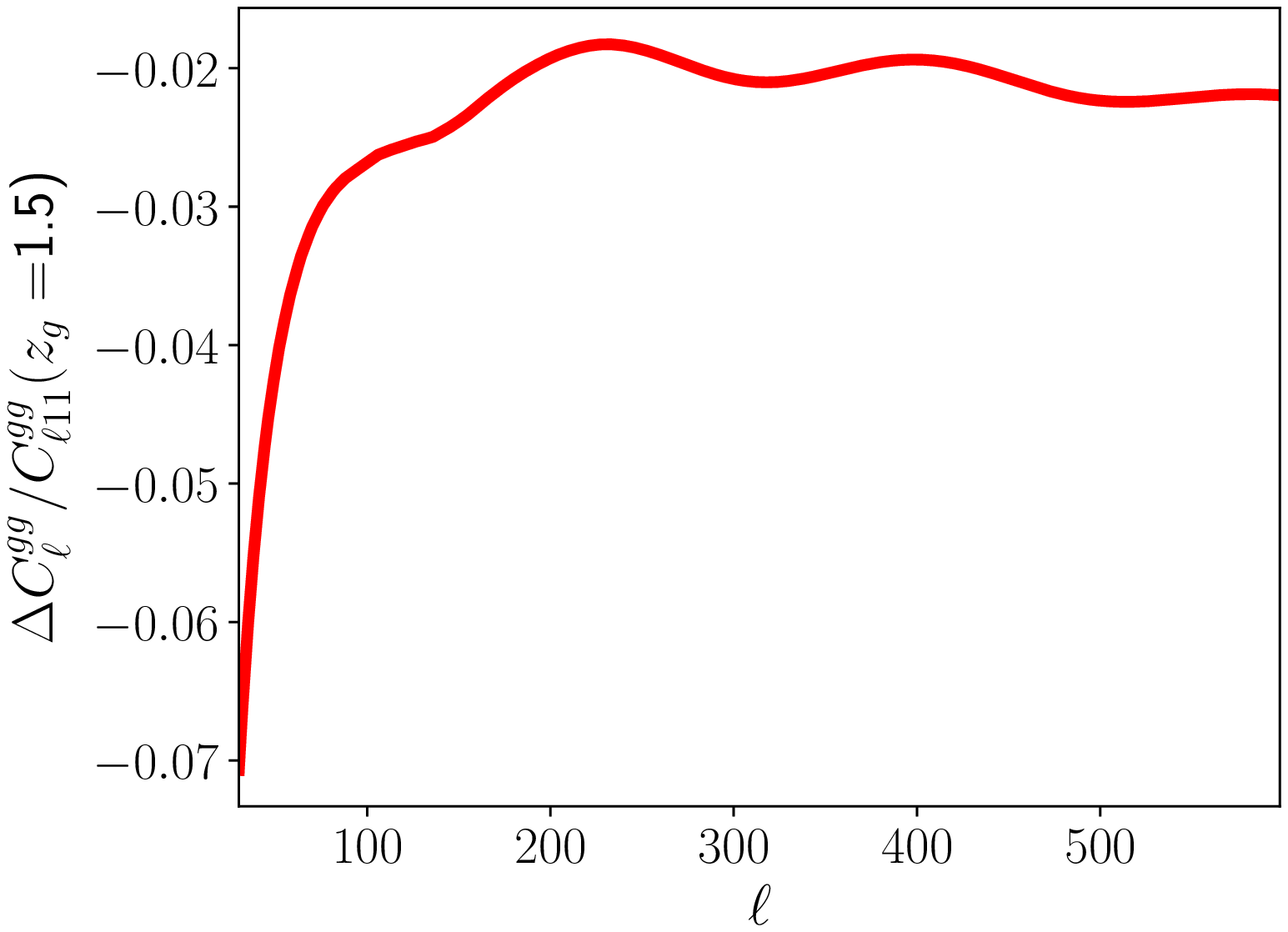}\\
	\caption{Same as Figure \ref{fig:1} except for the $C_l^{gg}$ deviation.}
\end{figure*}\par

\begin{figure}
	\centering
		\includegraphics[width=0.45\textwidth]{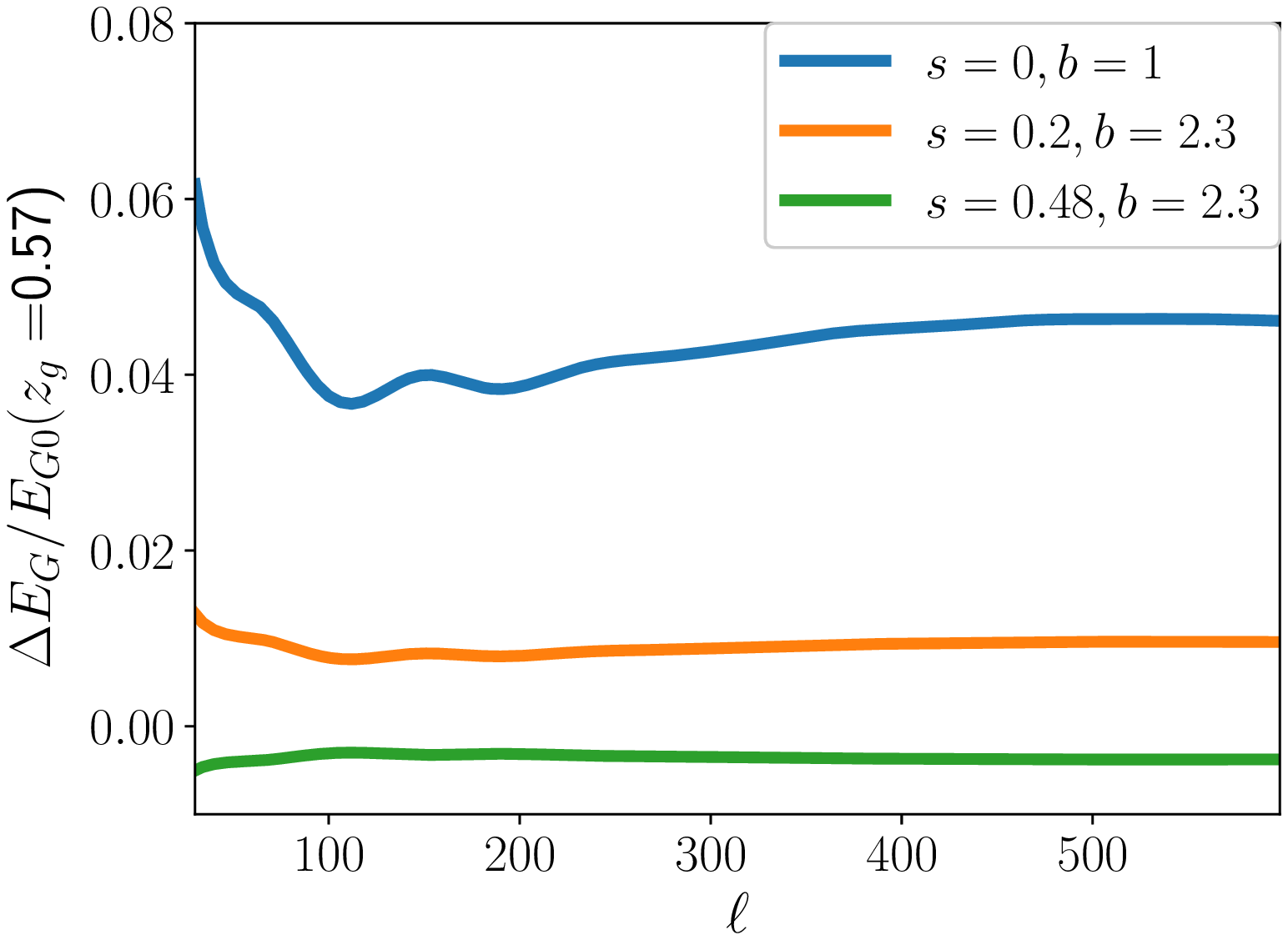}
		\includegraphics[width=0.45\textwidth]{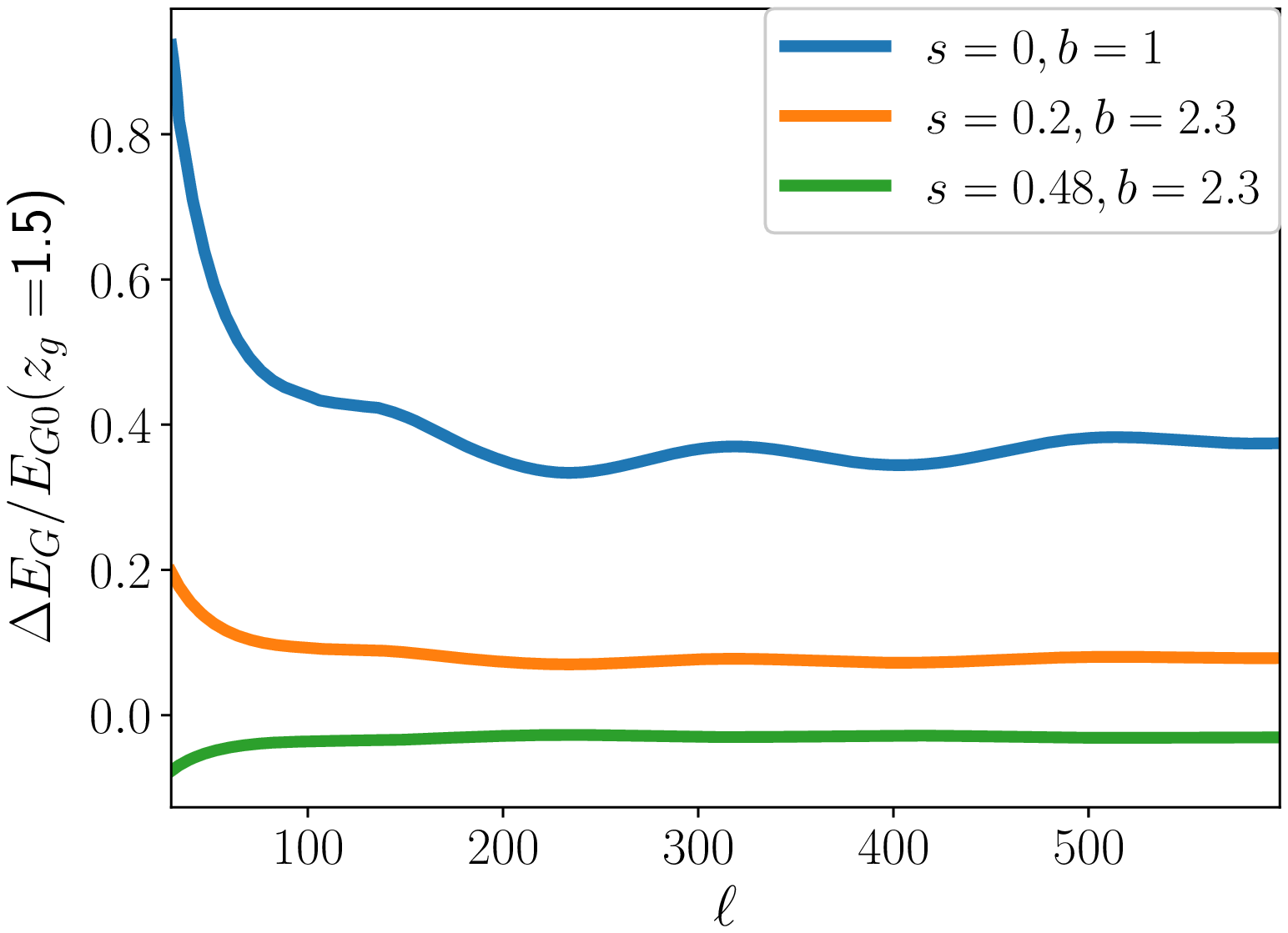}
	\caption{$E_G$ fractional deviation at redshifts centered at $z=0.57$ (top) and $z=1.5$ (bottom). $E_G$ is significantly biased when the galaxy bias and magnification bias parameter are set as $b=1$ and $s=0$, while for more realistic values, the $E_G$ bias is much smaller.}
	\label{fig:2}
\end{figure}

The deviation in $E_G$ due to magnification bias is shown in Figure \ref{fig:2}.
Using the fiducial values $s=0$ and $b=1$, we see significant deviations $\langle\Delta E_G/E_{G0}(\bar{z}=0.57)\rangle=4.37\%$ and $\langle\Delta E_G/E_{G0}(\bar{z}=1.5)\rangle=39.35\%$. However, more common values for the magnification bias for upcoming surveys are $s=0.2$ and 0.48, as shown in Table \ref{t:1}. Setting $s=0.2$ and $b=2.3$ decreases the deviations significantly to $\langle\Delta E_G/E_{G0}(\bar{z}=0.57)\rangle=0.91\%$ and $\langle\Delta E_G/E_{G0}(\bar{z}=1.5)\rangle=8.27\%$. Setting $s=0.48$ and $b=2.3$ changes the deviations to $\langle\Delta E_G/E_{G0}(\bar{z}=0.57)\rangle=-0.36\%$ and $\langle\Delta E_G/E_{G0}(\bar{z}=1.5)\rangle=-3.23\%$.  %In order to forecast whether magnification bias will cause a problem in measurements, we estimate $E_G$ error using 100 simulations constructed for BOSS+eBOSS LRG $\times$ Adv.~ACT survey and DESI QSO $\times$ Adv.~ACT survey, detailed in section \ref{S:calitest}. The $E_G$ fractional error averaged over all $\ell$-bins is about $2.8\%$ for $z_g=0.57$ and $3.8\%$ for $z_g=1.5$, which is larger than the magnification bias. 
%This indicates that the magnification effect is unlikely to cause a significant problem for $E_G$ measurements from LRGs or QSOs at $z=0.57$ and $z=1.5$, assuming redshift distribution shown in Eq.~\ref{eq:20}.
This indicates that the magnification effect is unlikely to cause a significant problem for $E_G$ measurements from galaxy surveys with large clustering bias and magnification factor $s$ close to 0.4. We do expect that for emission line galaxies, which tend to have lower clustering biases, the magnification bias effect will be larger and potentially significant. We will test $E_G$ deviations under various galaxy and CMB survey cross-correlations using realistic redshift ranges and galaxy clustering and magnification biases in section \ref{S:calitest}.\par

\subsection{Calibration method for magnification bias} \label{S:cali}

The true $E_G$ should be estimated using $C_\ell^{\kappa g1}$ and $C_\ell^{g1g1}$ which are not contaminated by magnification bias. Since we cannot estimate these directly from data, we have to instead find a way to subtract the higher order terms in Eq. \ref{eq:17} and Eq. \ref{eq:18} for each $C_\ell$ from magnification bias.  Since we are testing cosmology, we cannot just compute them theoretically; however, by looking at the expressions in Eqs.~\ref{E:gg} and \ref{E:kg}, we see that $C_\ell^{\kappa\kappa}$, given by Eq.~\ref{E:clab} for $A=B=\kappa$, shares a very similar form with $C_\ell^{\kappa g2}$ and $C_\ell^{g2g2}$ while $C_\ell^{\kappa g1}$ has a similar form to $C_\ell^{g1g2}$.  This implies that these spectra should be highly correlated, such that we can use the measured $C_\ell^{\kappa\kappa}$ to predict the correlations to $C_\ell^{\kappa g}$ and $C_\ell^{gg}$.

In order to predict these correlations, we propose the following scaling formulae:
\begin{eqnarray}\label{eq:30}
    \tilde{C}_\ell^{\kappa g2}=\left(\frac{C_\ell^{\kappa g2}}{C_\ell^{\kappa\kappa}}\right)\hat{C}_\ell^{\kappa\kappa}
\end{eqnarray}
\begin{eqnarray}\label{eq:31}
    \tilde{C}_\ell^{g2g2}=\left(\frac{C_\ell^{g2g2}}{C_\ell^{\kappa\kappa}}\right)\hat{C}_\ell^{\kappa\kappa}
\end{eqnarray}
\begin{eqnarray}\label{eq:32}
    \tilde{C}_\ell^{g1g2}&=&\left(\frac{C_\ell^{g1g2}}{C_\ell^{\kappa g1}}\right)\tilde{C}_\ell^{\kappa g1}\nonumber\\
    &=&\left(\frac{C_\ell^{g1g2}}{C_\ell^{\kappa g1}}\right)\times\left[\hat{C}_\ell^{\kappa g}-\left(\frac{C_\ell^{\kappa g2}}{C_\ell^{\kappa\kappa}}\right)\hat{C}_\ell^{\kappa\kappa}\right]
\end{eqnarray}
Here we use hats to denote quantities estimated from data.  We use tildes to denote the scaled correlations. Quantities without a marker are computed from theory.  These estimated terms will then be subtracted from measured $\hat{C}_\ell^{\kappa g}$ and $\hat{C}_\ell^{gg}$ to estimate $C_\ell^{\kappa g1}$ and $C_\ell^{g1g1}$ to compute $E_G$.  In principle, there should be better ways to reduce magnification bias.  For example, we could instead perform an Monte Carlo Markov Chain (MCMC) analysis where we fit for $E_G$ and the magnification bias terms in the $C_\ell$s simultaneously.  However, the goal of our analysis is to find the simplest way to calibrate magnification bias.  The next section uses simulations to determine how well these calibrations work.  Also, we expect these correlations to vary slightly with cosmological parameters, which must be set to construct simulations to compute the correlations.  Thus, we also test how sensitive the calibrations are to the assumed cosmologies, and specifically how much bias is introduced by using cosmological parameters that are not the true values but are still within known uncertainties.   %These correlations allow us to further calibrate $E_G$ as:
%\begin{widetext}
%\begin{eqnarray}\label{eq:33}
   % E_G(l)=\frac{c^2}{3H_0^2}\frac{C_l^{mg}}{Q_l^{mg}}C_b(l)\frac{\hat{C}^{\kappa g}_l-(C^{\kappa g}_{l2}/C^{\kappa\kappa}_l)\hat{C}^{\kappa\kappa}_l}{\beta[\hat{C}^{gg}_l-2(C^{gg}_{l12}/C^{\kappa g}_{l1})(\hat{C}^{\kappa g}_l-(C^{\kappa g}_{l2}/C^{\kappa\kappa}_l)\hat{C}^{\kappa\kappa}_l)-(C^{gg}_{l12}/C^{\kappa\kappa}_l)\hat{C}^{\kappa\kappa}_l]}
%\end{eqnarray}
%\end{widetext}

\section{Results}\label{S:results}
In this section we compute the deviation of $E_G$ due to magnification bias for multiple spectroscopic and photometric surveys cross-correlated with CMB lensing maps from Planck and Adv.~ACT.  We find that most of the cases where there are significant deviations are from photometric surveys, although DESI can also exhibit a deviation.  We also test our calibration method, finding that it is efficient in removing magnification bias and it is stable with regard to small deviations in the parameters used to construct the calibration.

\subsection{Testing $E_G$ calibrations using simulations} \label{S:calitest}
We use theoretical angular power spectra of the CMB lensing convergence and galaxy overdensity to construct simulated maps.  We use these maps for two purposes.  The first purpose is to construct errors on $E_G$.  $E_G$, being a ratio between 3 measured quantities, is inherently non-Gaussian. While we can use error propagation to get a relatively accurate estimate of the statistical error on $E_G$, simulations can naturally capture the additional contribution to the statistical fluctuations due to this non-Gaussianity.  The second purpose of these simulations is to test our proposed calibration method to remove significant magnification bias.  In particular, we expect the measured $C_\ell^{\kappa g}$ and calibration terms scaled to $C_\ell^{\kappa\kappa}$ to be correlated, and simulations allow us to capture this effect naturally.

The pipeline of the simulation process we use is described in \cite{Serra2014}. In short, we compute all the correlation terms listed in Eqs.~\ref{eq:24}-\ref{eq:29} and follow Eqs.~\ref{eq:17}-\ref{eq:18} to obtain theoretical values of observables $C_\ell^{gg}$, $C_\ell^{\kappa g}$ and $C_\ell^{\kappa\kappa}$. In each simulation, we construct the covariance matrix $C$ and decompose it using Cholesky decomposition as $C=LL^T$ with $L$, a lower triangular matrix, given by
\begin{eqnarray}\label{eq:34}
    C=\begin{pmatrix}C^{\kappa\kappa}_l+N^{\kappa}_l&C^{\kappa g}_l\\C^{\kappa g}_l&C^{gg}_l+N^{g}_l\end{pmatrix}=LL^T
\end{eqnarray}
Here $N^{\kappa}_l$ and $N^g_l$ are lensing noise and shot noise of CMB and galaxy surveys cross-correlated in the simulation.
Spherical harmonic coefficients are then obtained through
\begin{eqnarray}
\begin{pmatrix}a^{\kappa}_{lm}\\a^g_{lm}\end{pmatrix}=L\begin{pmatrix}\xi_1\\\xi_2\end{pmatrix}
\end{eqnarray}
Here $\xi_1$ and $\xi_2$ are arrays of random numbers with zero mean and unit variance.\par
We use $a_{lm}^{\kappa}$ and $a_{lm}^g$ to compute full sky galaxy maps and CMB lensing maps through functions provided by the HEALPix package \citep{Gorski2005}. We apply $N_{\rm side}=1024$ for the galaxy maps and $N_{\rm side}=2048$ for the lensing maps, after which we then add a mask onto the full sky maps. We assume Planck and Adv.~ACT survey overlap completely with galaxy surveys well except that Adv.~ACT only overlaps with 75\% of the DESI survey area. We correlate masked galaxy map and CMB lensing map with each other to get angular power spectra of the partial sky maps $D_\ell^{gg}$, $D_\ell^{\kappa g}$ and $D_\ell^{\kappa\kappa}$ and then estimate $\hat{C}^{gg}_\ell$, $\hat{C}^{\kappa g}_\ell$ and $\hat{C}^{\kappa\kappa}_\ell$ from simulated data using the naive prescription \citep{Hivon2002}
\begin{eqnarray}
\hat{C}_\ell=\frac{\langle D_\ell\rangle}{f_{\rm sky}\omega_2}
\end{eqnarray}
where $f_{\rm sky}\omega_i$ is the $i$-th momentum of the mask $W(\hat{n})$
\begin{eqnarray}
f_{sky}\omega_i=\frac{1}{4\pi}\int_{4\pi}d\hat{n}W^i(\hat{n})
\end{eqnarray}\par
Following \cite{Pullen2015}, we assume that scales dominated by nonlinear clustering will not be used to estimated $E_G$.  Thus, we set the maximum wavenumber comprising linear to quasi-linear scales to $k_{nl}$, where $k^3_{nl}P(k_{nl},\bar{z})/(2\pi^2)=1$ and set the maximum scale $\ell_{max}=\chi(\bar{z})k_{nl}$. Throughout this work we use the non-linear matter power spectrum calculated by CAMB (HALOFIT) \cite{2000ApJ...538..473L}, but removing nonlinear scales from the estimator do not significantly affect the results. We then bin the simulated angular power spectrum into 10 bins with $\Delta_\ell=\ell_{max}/10$, after which we remove the lowest and highest $\ell$-bins, leaving us with 8 bins.  Note that since the galaxy surveys each have different median redshifts, the relevant scales will vary for each survey.\par
In our analysis we do include scales that could be contaminated by systematic effects other than magnification bias and nonlinear clustering.  These effects include redshift space distortions and stellar contamination at large scales and lensing Gaussian and point-source bias at small scales \citep{Pullen2016}.  However, the main focus of this paper is on magnification bias, and we leave the mitigation of these other systematic effects to future work.\par

Since we are assessing the $E_G$ calibration using simulations, we consider the total error in the simulated $E_G$, $\sigma_{tot}$, given by 
\begin{equation}
\begin{split}
    \sigma_{tot}&\sim\sqrt{\sigma^2_{stat}+\sigma^2_{dev}+\sigma^2_{sim}}\\
    &=\sigma_{stat}\sqrt{1+\left(\dfrac{\sigma_{dev}}{\sigma_{stat}}\right)^2+\left(\dfrac{\sigma_{sim}}{\sigma_{stat}}\right)^2}\\
    &\sim\sigma_{stat}\sqrt{1+\left(\dfrac{\sigma_{dev}}{\sigma_{stat}}\right)^2+\dfrac{1}{N_{\rm sim}}}
\end{split}
\end{equation}
where $\sigma_{stat}$ is the statistical $E_G$ error, $\sigma_{dev}$ is the deviation of $E_G$ caused by magnification bias, and $\sigma_{sim}$ is the error due to using a finite number of simulations $N_{\rm sim}$, which decreases as $\sigma_{sim}\sim\dfrac{\sigma_{stat}}{\sqrt{N_{\rm sim}}}$.  Our goal is to find $E_G$ deviations that are at least 50\% of the statistical error, which would increase the total $E_G$ error by 10\%.  According to our expression, using 100 simulations should be enough such that the total error should not increase significantly; thus we choose to use 100 simulations in our analysis.
\par
The CMB surveys we consider are the Planck satellite and Adv.~ACT. It was shown recently that non-Gaussian clustering in the lenses can bias CMB lensing estimators dominated by the temperature maps at the 3\% level compared to the signal \citep{2018arXiv180601157B,2018arXiv180601216B}, while the bias can be as high as 7\% for cross-correlations with large-scale structure tracers \citep{2018arXiv180601157B}, though the latter prediction is still preliminary (B\"{o}hm et al. in prep).  These biases should affect CMB lensing estimates from both Planck and Adv.~ACT.  However, it was also shown in \citet{2018arXiv180601157B} that polarization-based estimates of CMB lensing do not have this bias.  Thus, although the sensitivity of CMB-S4 is not demonstrably better than Adv.~ACT, its lensing estimate would be more accurate for an $E_G$ analysis since its polarization maps would dominate its CMB lensing estimate.\par
The galaxy surveys include both spectroscopic and photometric surveys listed in Table \ref{t:1}.  Since estimating $E_G$ requires a measurement of $\beta$, photometric galaxy surveys would normally not be considered because the lower redshift precision washes out the RSD effect.  However, as argued in \citet{Pullen2015}, the larger number of galaxies make photometric surveys so much more precise in measuring $C_\ell^{\kappa g}$ that although the precision in $\beta$ from a photometric survey is lower \citep{2011MNRAS.415.2193R,2014MNRAS.445.2825A}, the overall error in $E_G$ could still be lower generally than in spectroscopic surveys, particularly if catastrophic redshift errors are kept to a minimum.
\par In all simulations in this subsection we assume a $\Lambda$CDM cosmogony. We assume an eBOSS LRG redshift distribution similar to \cite{2016AJ....151...44D} Table 1 but in the range $z\in[0.4,0.9]$. The eBOSS QSO redshift distribution we use is similar to that in \cite{Parisa} but in the redshift range $z\in[0.9,3.5]$, and we use an expression for the QSO clustering bias from Table 2 in \cite{Zhao2016}. The redshift distributions for DESI survey are from \cite{DESICollaboration2016} Table 2.5.  For the Euclid spectroscopic survey we use a redshift distribution estimated from the H$\alpha$ luminosity function from \citet{2013ApJ...779...34C} with a flux limit of $4\times 10^{-16}$ erg/s/cm$^2$. For both DES and LSST, we use the toy redshift distribution model from \citet{2014JCAP...05..023F}, given by
\begin{eqnarray}
f_g(z)=\frac{\eta}{\Gamma\left(\frac{\alpha+1}{\eta}\right)z_0}\left(\frac{z}{z_0}\right)^\alpha\exp^{-(z/z_0)^\eta}\, ,
\end{eqnarray}
where the parameter values are given by $(\alpha,\eta,z_0)=(1.25,2.29,0.88)$ for DES and $(2.0,1.0,0.3)$ for LSST.  For the Euclid photometric survey we apply estimates of redshift distributions based on \cite{Hsu2014a,Guo2013}.\par
It can be shown that $s=-(1+\alpha_{LF})/2.5$, where $s$ is the magnification bias factor we define in Eq. \ref{eq:13}, and $\alpha_{LF}$ is the faint-end slope parameter in the Schechter luminosity function \citep{1976ApJ...203..297S}. We present the proof in Appendix \ref{S:mag_fac}. Our $s$ factor sources for BOSS LRG, eBOSS ELG, BOSS QSO, DESI (LRG, ELG, QSO) and Euclid spectroscopic surveys are in sequence \cite{2012ApJ...746..138T}, \cite{Comparat:2014xza}, \cite{2013ApJ...773...14R}, \cite{Aghamousa:2016zmz}, and \cite{2013ApJ...779...34C}. We use the $s$ factor from BOSS or eBOSS survey as BOSS+eBOSS combined magnification bias factor. We use (A.5) in \cite{Montanari:2015rga} to estimate the $s$ factor for all the three photometric surveys. \par
Since the eBOSS/BOSS areas overlap for both the LRGs and the QSOs, we use the eBOSS area as the total area. We use $\bar{N}$ to denote galaxy number density per steradian, and we estimate galaxy shot noise for $C_\ell^{gg}$ as $N_\ell^g=1/\bar{N}$. We assume $\bar{N}_{BOSS+eBOSS}=\bar{N}_{BOSS}+\bar{N}_{eBOSS}$ for BOSS$+$eBOSS survey. Since the galaxy bias we use here is no longer a constant, we use $C_b=C^{g1g1}_{\ell}/[b(\bar{z})C^{mg}_\ell]$ to calibrate $E_G$ bias due to galaxy bias evolution \citep{Reyes2010a,Pullen2016}. Details we apply for different galaxy surveys including $k_{nl}$, $\ell_{max}$ and biases $s$ and $b(z)$ are summarized in Table \ref{t:1}.
\begin{table*}
    \centering
    \begin{tabular}{l c c c c c c c}
    \hline\hline        &$b(z)$&s&z&Area($deg^2$)&$N_{gal}$&$k_{nl}(h/Mpc)$&$l_{max}$  \\\hline
         BOSS LRG&$1.7/D(z)$&&0.43-0.7&7900&704,000&0.275&380\\\hline
         eBOSS LRG&$1.7/D(z)$&&0.6-0.8&7500&300,000&0.299&500\\\hline
         BOSS+eBOSS LRG&$1.7/D(z)$&0.2&0.43-0.8&7500&968,000&0.288&450\\\hline
         eBOSS ELG&$1/D(z)$&0.48&0.6-1.0&1000&189,000&0.318&620\\\hline
         BOSS QSO&$0.53+0.29(1+z)^2$&&2.1-3.5&10,200&175,000&0.611&1000\\\hline
         eBOSS QSO&$0.53+0.29(1+z)^2$&&0.9-3.5&7500&573,000&0.450&1000\\\hline
         BOSS+eBOSS QSO&$0.53+0.29(1+z)^2$&0.2&0.9-3.5&7500&701,000&0.476&1000\\\hline
         DESI LRG&$1.7/D(z)$&0.87&0.6-1.2&14,000&$4.1\times10^6$&0.312&580\\\hline
         DESI ELG&$0.84/D(z)$&0.1&0.6-1.7&14,000&$1.8\times10^7$&0.352&810\\\hline
         DESI QSO&$1.2/D(z)$&0.29&0.6-1.9&14,000&$1.9\times10^6$&0.400&1000\\\hline
         Euclid (spectro)&$0.9+0.4z$&0.16
         &0.5-2.0&20,000&$3.0\times10^7$&0.326&660\\\hline
         DES&$0.9+0.4z$&0.29&0.0-2.0&5,000&$2.16\times10^8$&0.312&580\\\hline
         LSST&$0.9+0.4z$&0.324&0.0-2.5&20,000&$3.6\times10^9$&0.330&680\\\hline
         Euclid (photo)&$0.9+0.4z$&0.326&0.0-3.7&20,000&$1.86\times10^9$&0.330&690\\\hline
    \end{tabular}
    \caption{Properties of spectroscopic and photometric surveys in our analysis.}
\label{t:1}
\end{table*}
\par
The standard deviation of multiple simulations give error estimation about $\hat{C}^{gg}_\ell$, $\hat{C}^{\kappa g}_\ell$ and $\hat{C}^{\kappa\kappa}_\ell$ as well as their covariance. We estimate $E_G$ errors following error propagation as
\begin{eqnarray}
    \frac{\sigma^2(\hat{E}_g)}{\hat{E}_g^2}&=&\frac{\sigma^2(\hat{C}_\ell^{\kappa g})}{(\hat{C}_\ell^{\kappa g})^2}+\frac{\sigma^2(\beta)}{\beta^2}+\frac{\sigma^2(\hat{C}_\ell^{gg})}{(\hat{C}_\ell^{gg})^2}\nonumber\\
    &&-2Cov(\beta,\hat{C}_\ell^{\kappa g})\frac{1}{\hat{C}_\ell^{\kappa g}\beta}\nonumber+2Cov(\beta,\hat{C}_\ell^{gg})\frac{1}{\beta \hat{C}_\ell^{gg}}\nonumber\\
    &&-2Cov(\hat{C}_\ell^{\kappa g},\hat{C}_\ell^{gg})\frac{1}{\hat{C}_\ell^{\kappa g}\hat{C}_\ell^{gg}}
\end{eqnarray}
In our analysis we ignore the term $-2Cov(\beta,\hat{C}_\ell^{\kappa g})/[\hat{C}_\ell^{\kappa g}\beta]$ since the CMB lensing convergence field, unlike the galaxy overdensity, includes perturbations from LSS at all redshifts. $\hat{C}_\ell^{\kappa g}$ therefore should be much less correlated to $\beta$ than $\hat{C}^{gg}_\ell$. In order to compute the fifth term, we apply $Cov(\beta,\hat{C}_\ell^{gg})=r\sigma(\hat{C}_\ell^{gg})\sigma(\beta)$ where the correlation coefficient $r$ is computed using simulations of the galaxy distribution in redshift space implemented in \cite{Pullen2016}. We check that the $2r\sigma(\Hat{C}_\ell^{gg})\sigma(\beta)/(\beta \hat{C}^{gg}_\ell)$ term only contributes less than $8\%$ of $\sigma(\hat{E}_g)$ in all cases we consider, so we also ignore this term.  A more precise analysis would use full 3D simulations along light cones in order to simulate CMB lensing, galaxy clustering, and RSD simultaneously; however, we save this for future work.\par
For spectroscopic surveys, we use error predictions of the $\beta$ forecasted by the surveys themselves. We use $\sigma(\beta)=0.025\beta$, $\sigma(\beta)=0.029\beta$ and $\sigma(\beta)=0.035\beta$ for the whole range for BOSS+eBOSS LRG, BOSS+eBOSS QSO, and eBOSS ELG, respectively \citep{2016AJ....151...44D,Zhao2016}; 4\% error in $\beta$ within $\Delta z=0.1$ for the DESI survey \citep{DESICollaboration2016}, and 3\% error in $\beta$ within $\Delta z=0.1$ for the Euclid spectroscopic survey \citep{Amendola2012}.  For photometric surveys, we assume errors for $\beta$ based on work by \citet{2011MNRAS.415.2193R} and \citet{2014MNRAS.445.2825A}. For DES, we use $\sigma(\beta)/\beta=0.17\sqrt{0.1(1+1)/(z_2-z_1)}$, while for LSST and photometric Euclid we expect a volume 4 times larger and thus $\sigma(\beta)/\beta=0.085\sqrt{0.1(1+1)/(z_2-z_1)}$, where $z_2$ and $z_1$ are the upper and lower limit of the redshift distribution.\par
%\Ant{which cases? For all galaxy survey $\times$ CMB survey. The largest ratio 7.46\% is from DESIQSO $\times$ Planck 'no noise'}

\begin{figure*}
\begin{center}
\begin{tabular}{ccc}	    &\Large{Planck}&\Large{Adv.~ACT}\\%&CMB-S4\\
		\rotatebox[origin=c]{90}{\Large{DESI LRG}}&\raisebox{-0.5\height}{\includegraphics[width = 7cm, height=6cm]{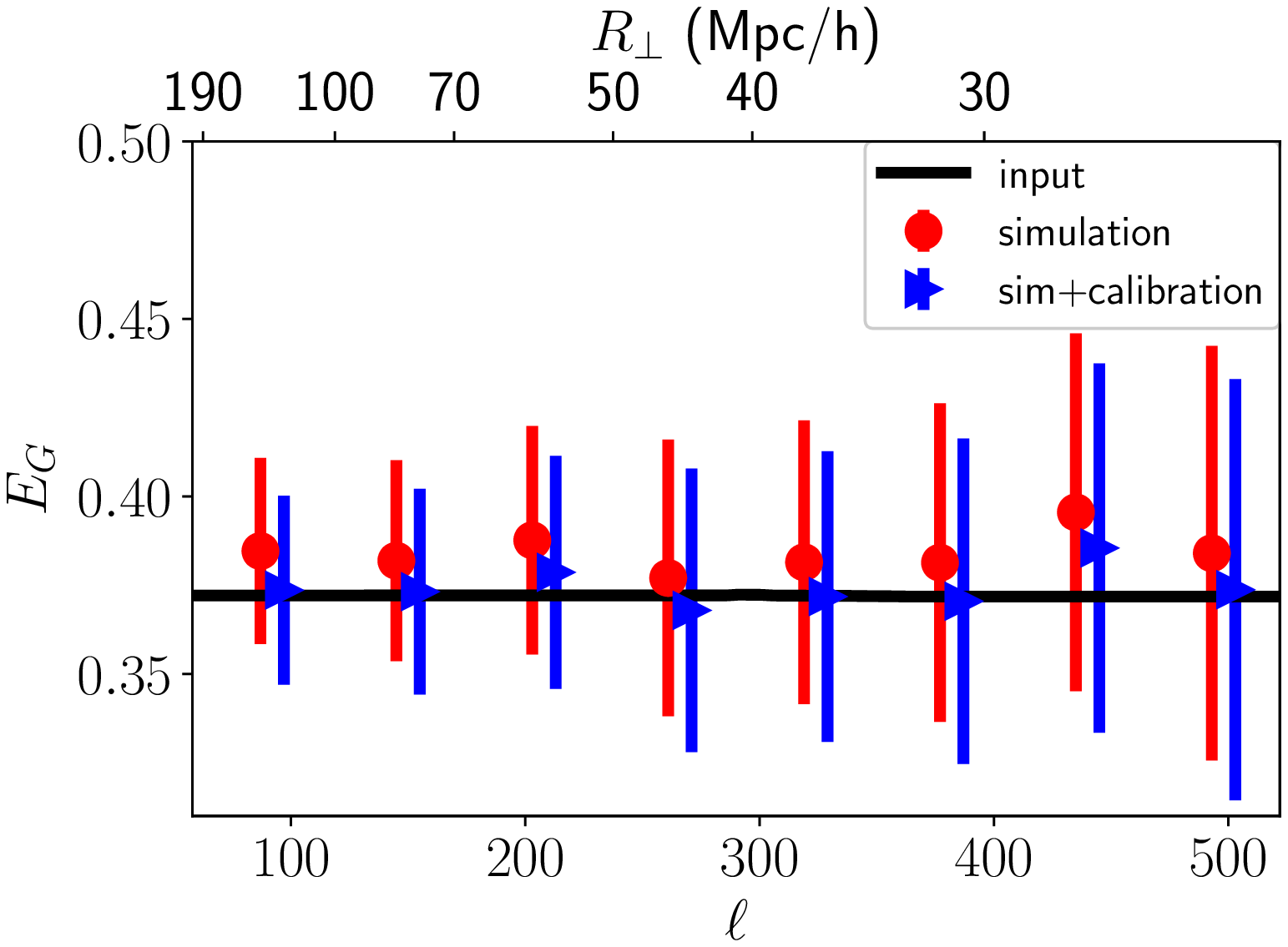}}&\raisebox{-0.5\height}{
		\includegraphics[width = 7cm, height=6cm]{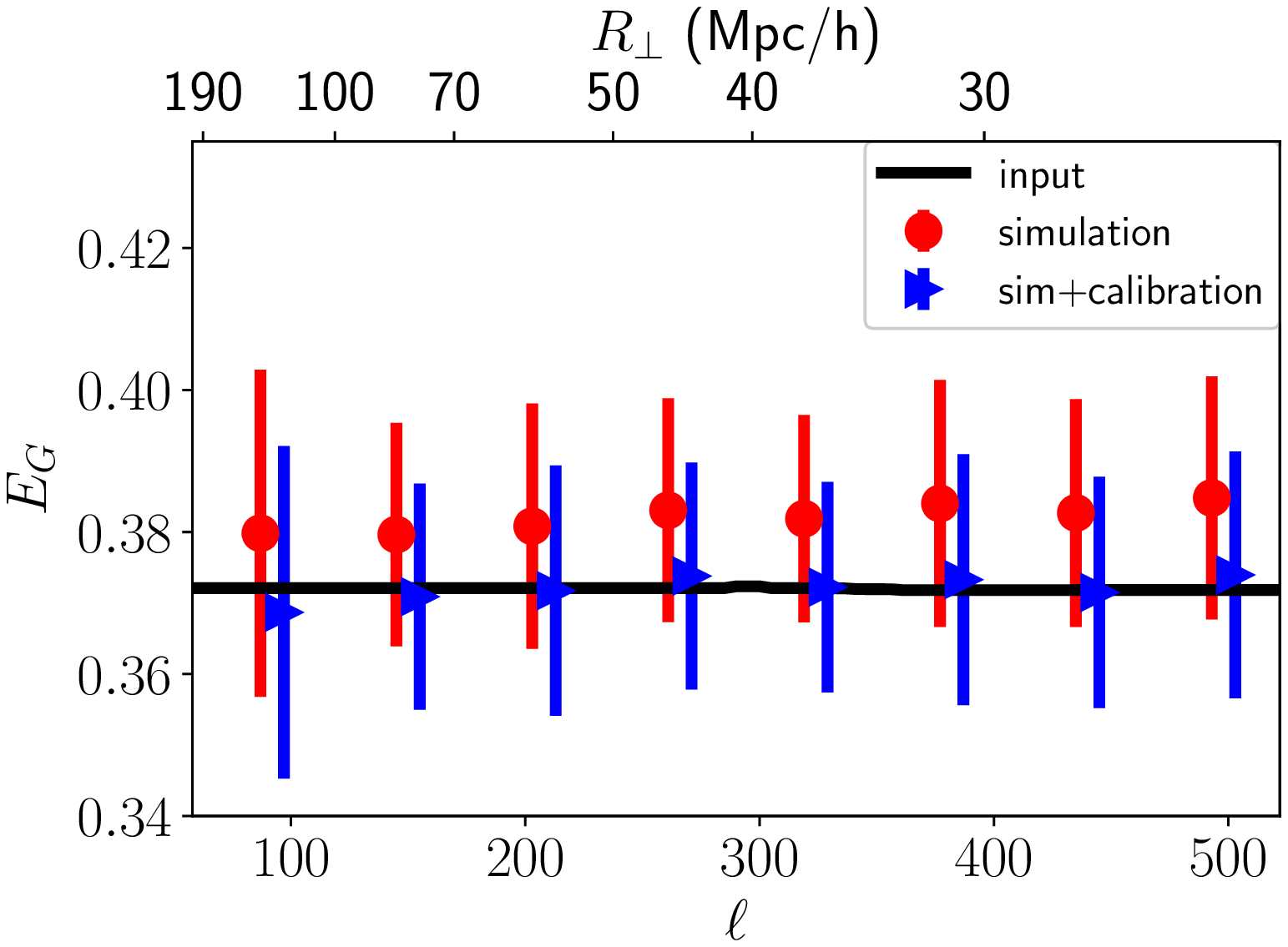}}\\
		\rotatebox[origin=c]{90}{\Large{DESI ELG}}&\raisebox{-0.5\height}{\includegraphics[width = 7cm, height=6cm]{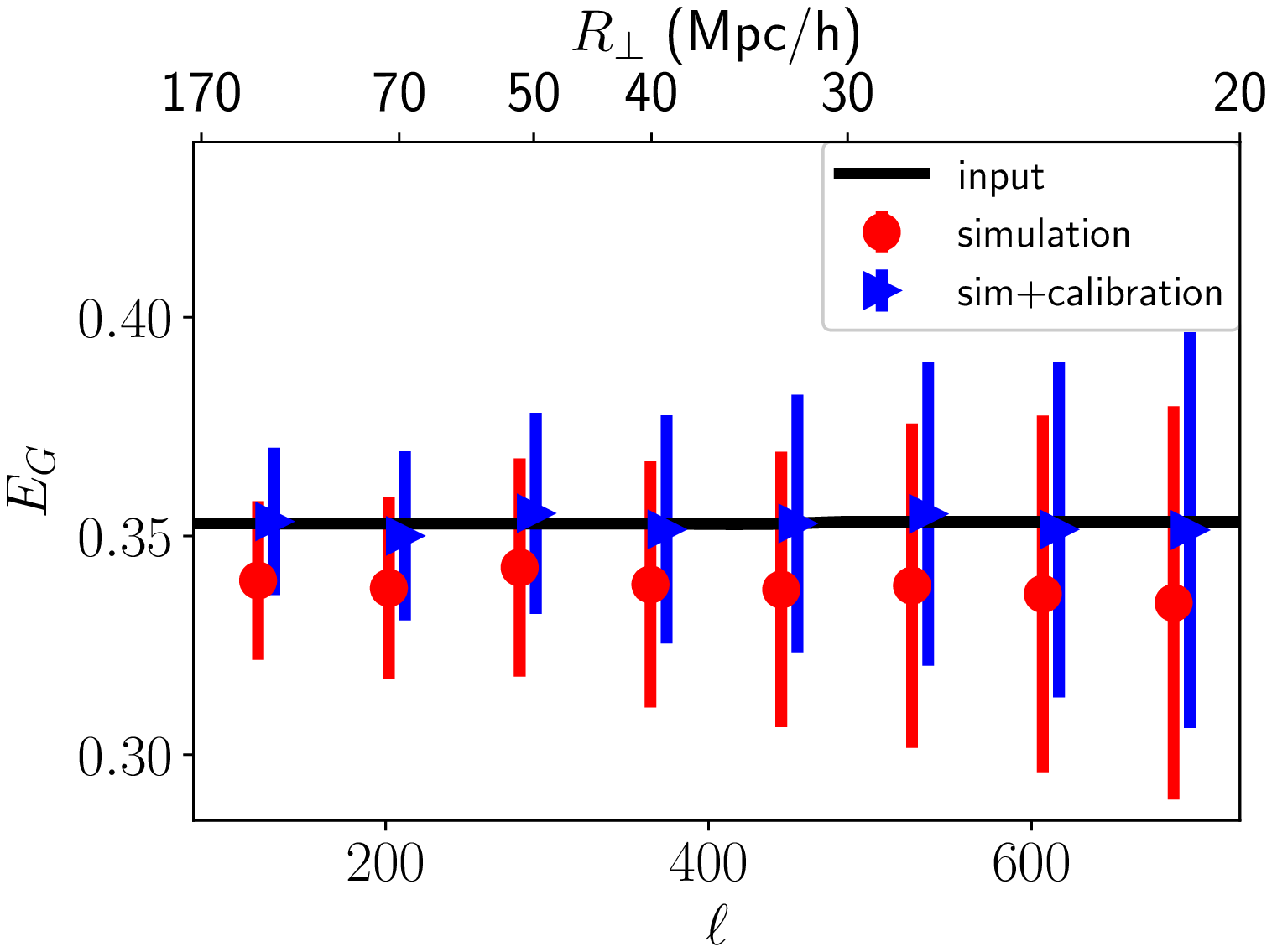}}&\raisebox{-0.5\height}{
		\includegraphics[width = 7cm, height=6cm]{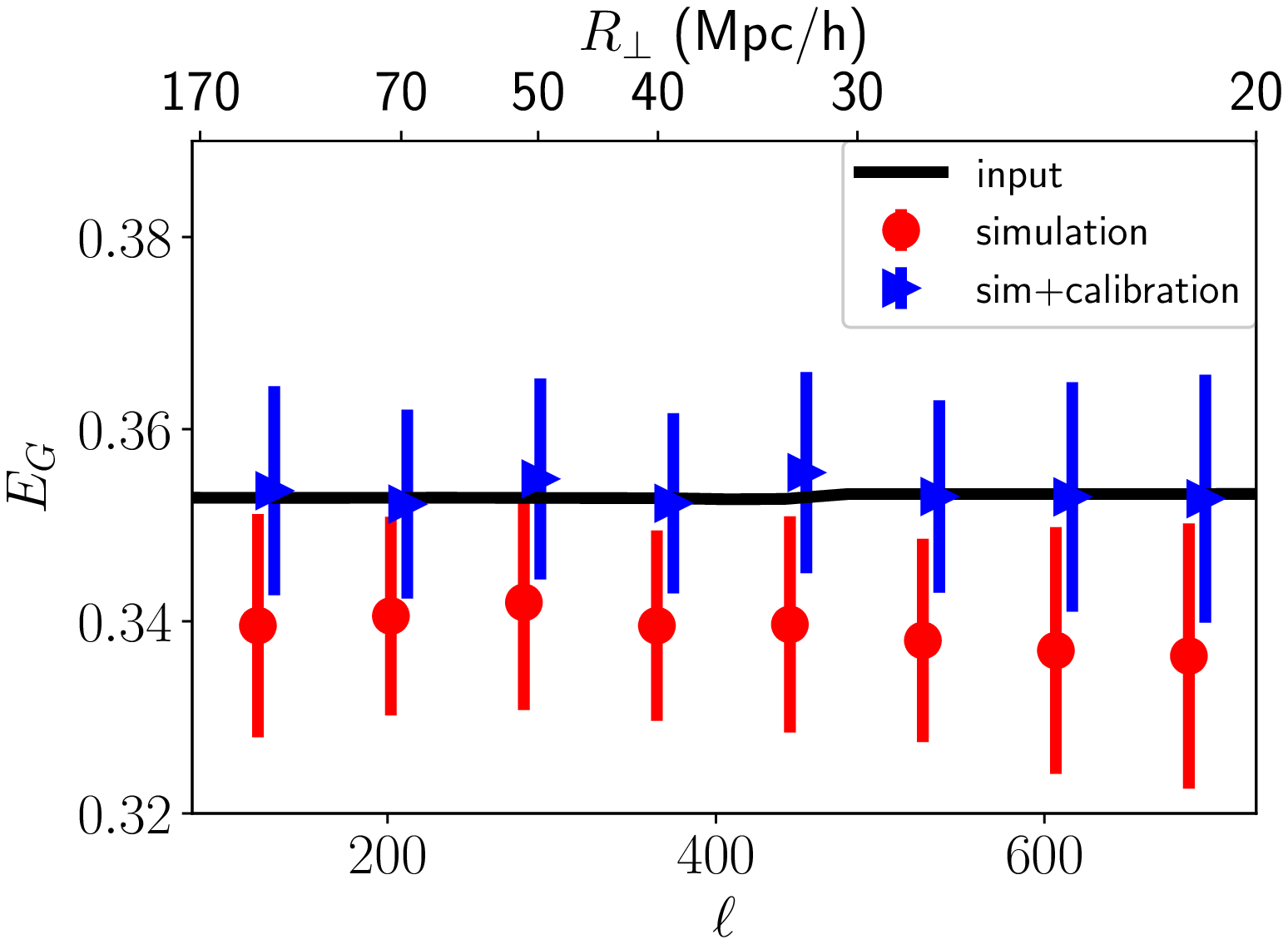}}\\
\end{tabular}		
\end{center}
\caption{Simulated $E_G$ measurements for spectroscopic surveys with and without magnification bias calibration.  The first column includes CMB lensing noise from the Planck survey while the second column assumes Adv. ACTPol. The top row is for DESI LRG, while the bottom row is for DESI ELG. We shift data after calibration by $\Delta l=10$ for reading convenience.}
\label{fig:3}
\end{figure*}

\begin{figure*}
\begin{center}
\begin{tabular}{ccc}	    
        &\Large{Planck}&\Large{Adv.~ACT}\\%&CMB-S4\\
		\rotatebox[origin=c]{90}{\Large{DES}}&\raisebox{-0.5\height}{\includegraphics[width = 7cm, height=6cm]{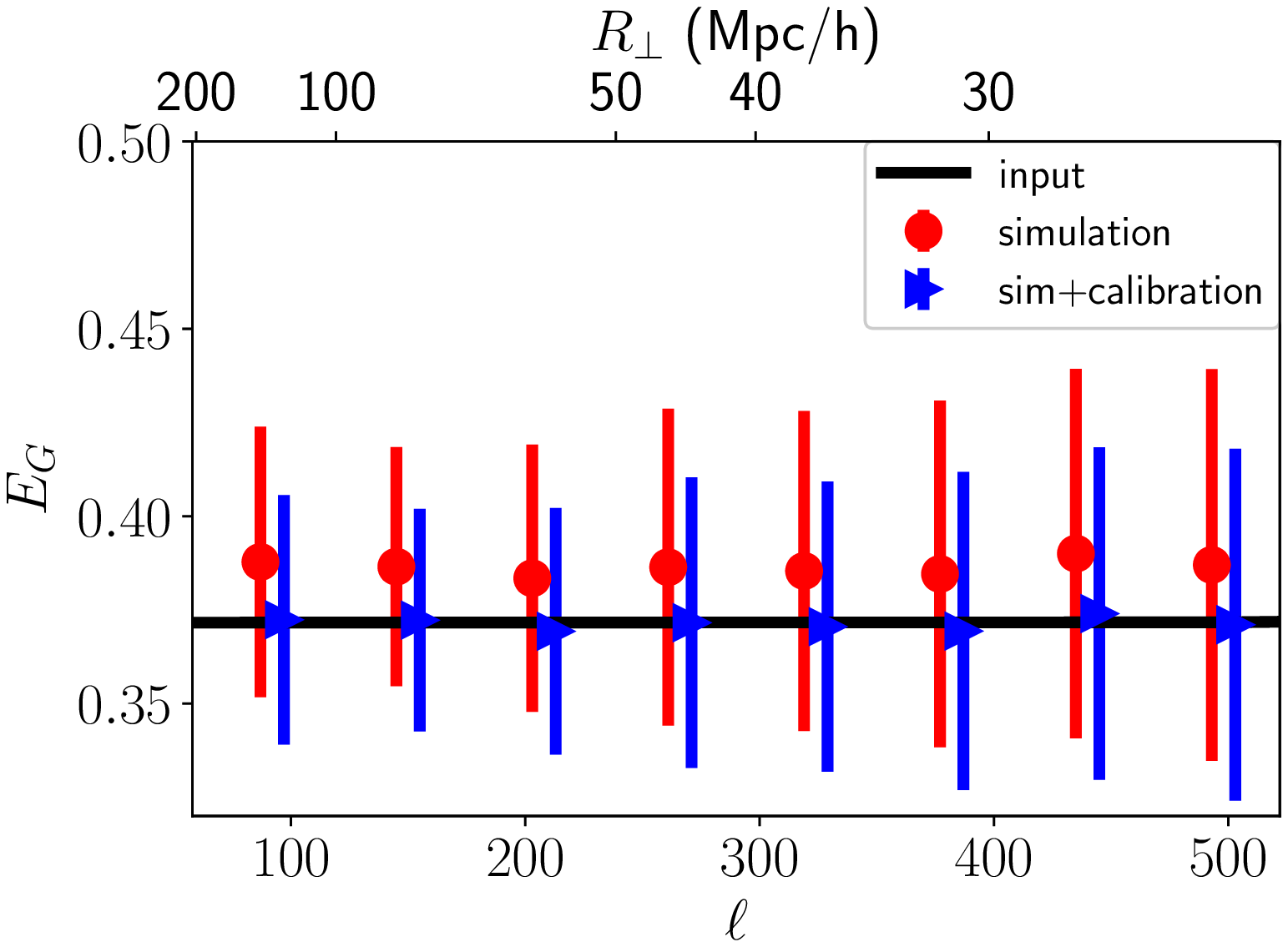}}&
		\raisebox{-0.5\height}{\includegraphics[width = 7cm, height=6cm]{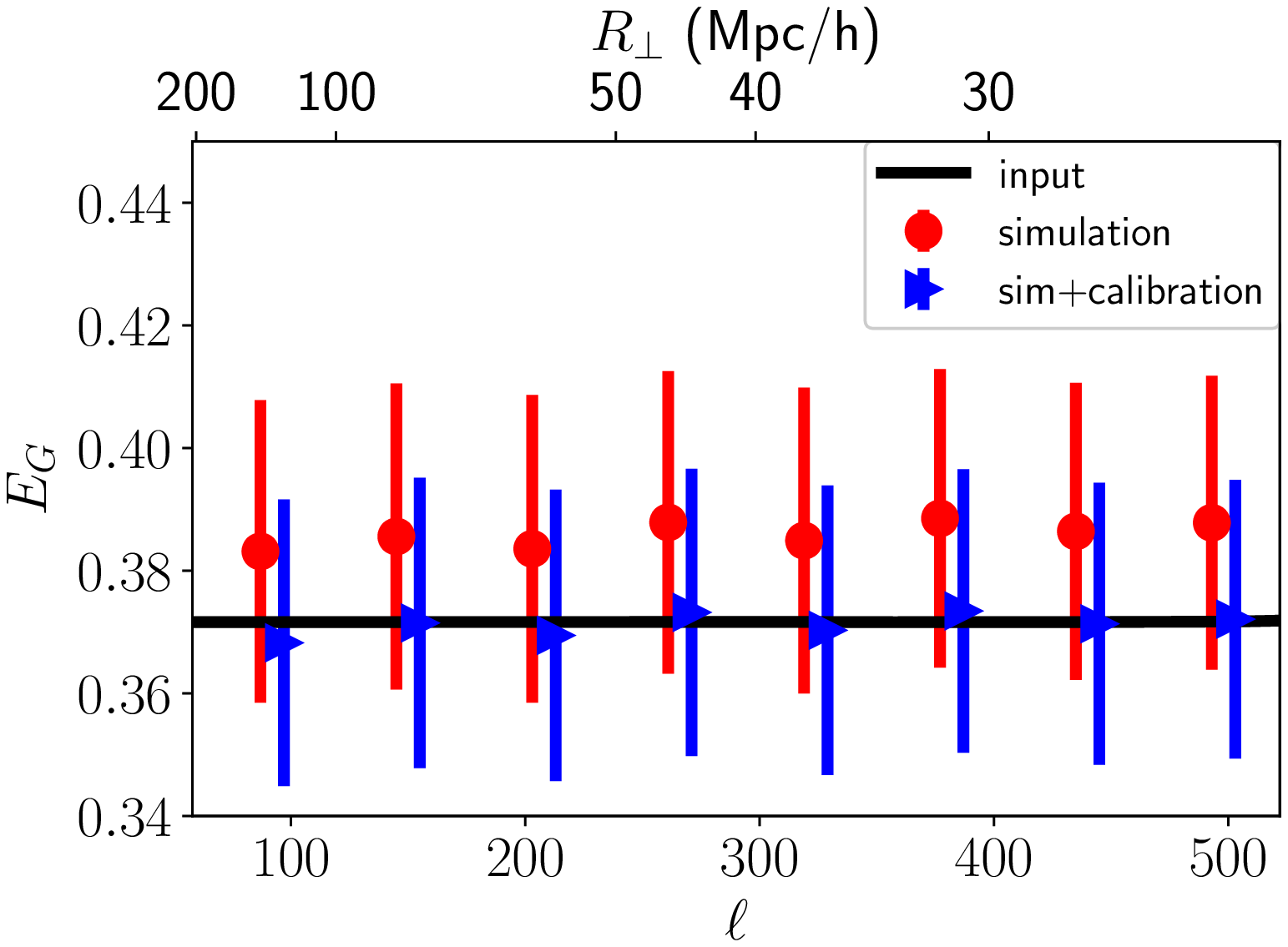}}\\
		\rotatebox[origin=c]{90}{\Large{LSST}}&\raisebox{-0.5\height}{\includegraphics[width = 7cm, height=6cm]{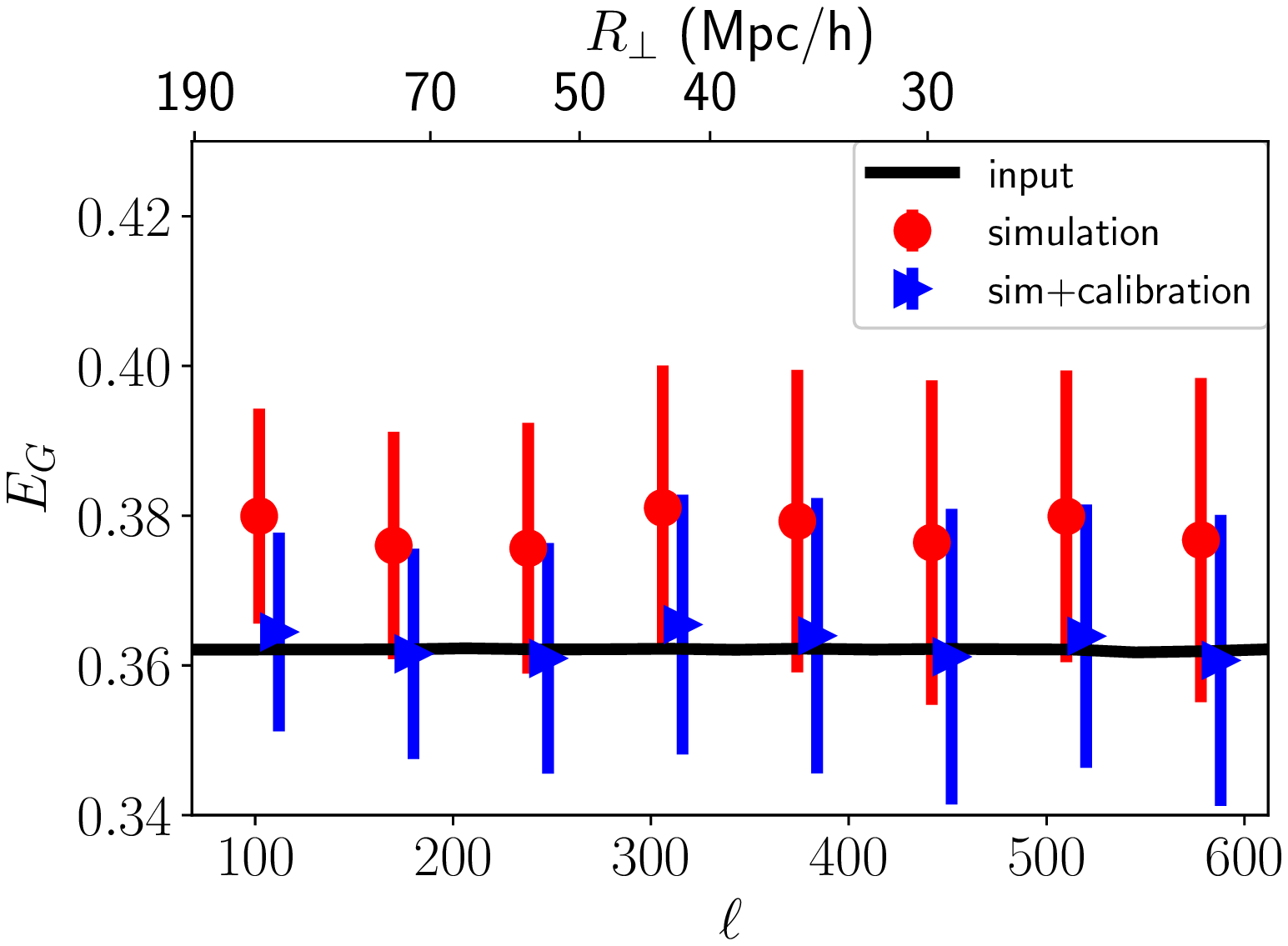}}&
		\raisebox{-0.5\height}{\includegraphics[width = 7cm, height=6cm]{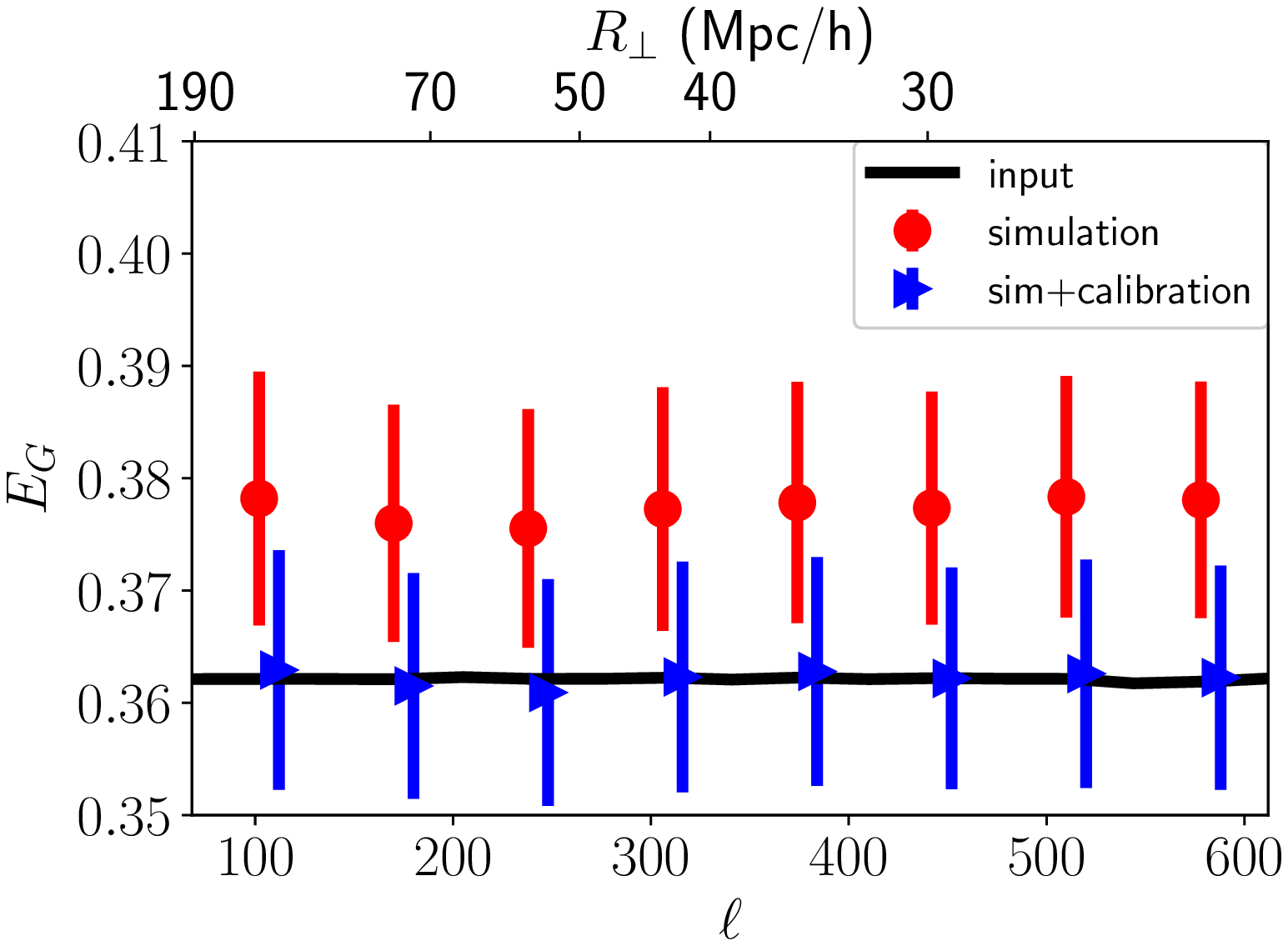}}\\
		\rotatebox[origin=c]{90}{\Large{Euclid (photo)}}&\raisebox{-0.5\height}{\includegraphics[width = 7cm, height=6cm]{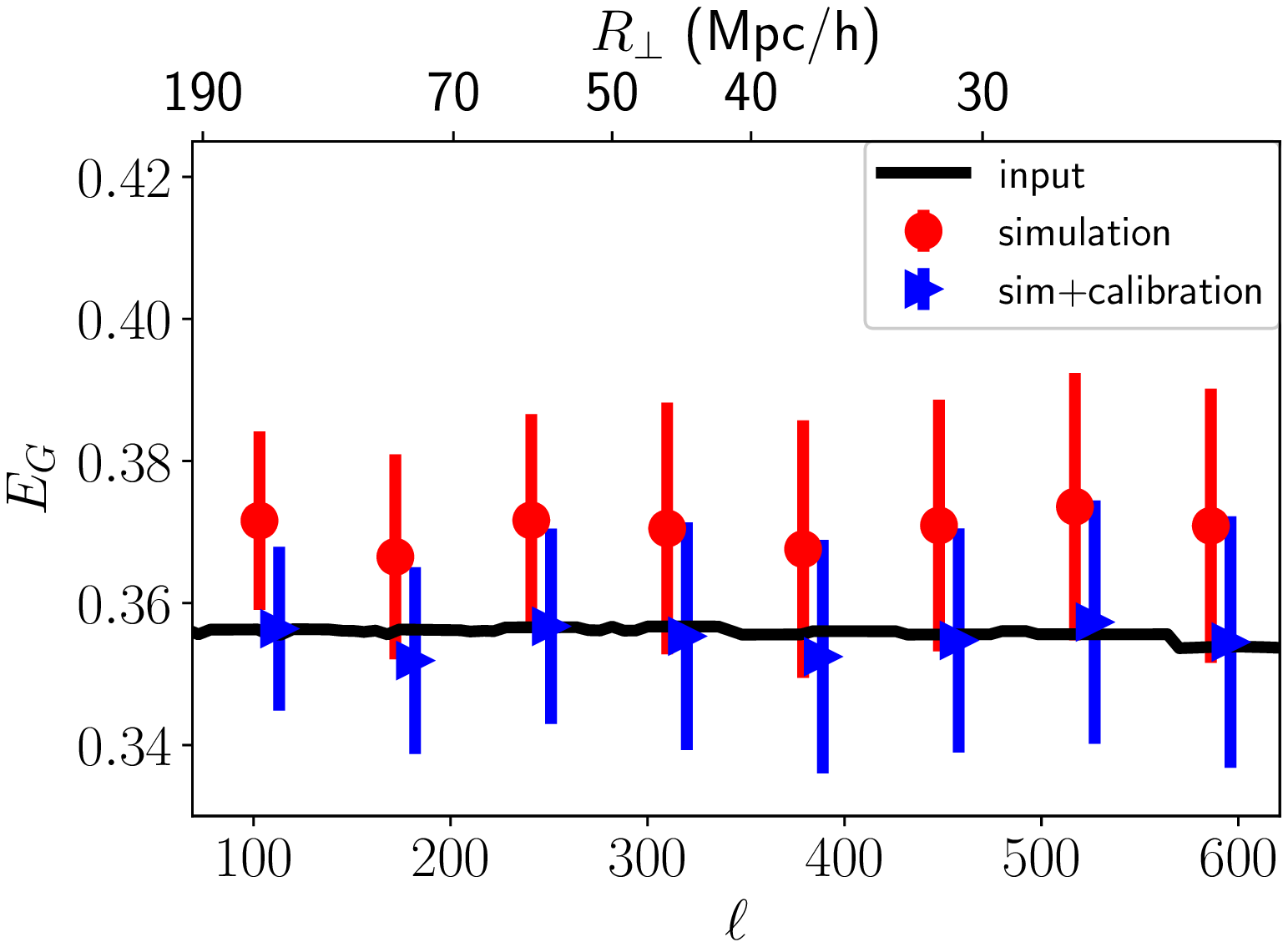}}&
		\raisebox{-0.5\height}{\includegraphics[width = 7cm, height=6cm]{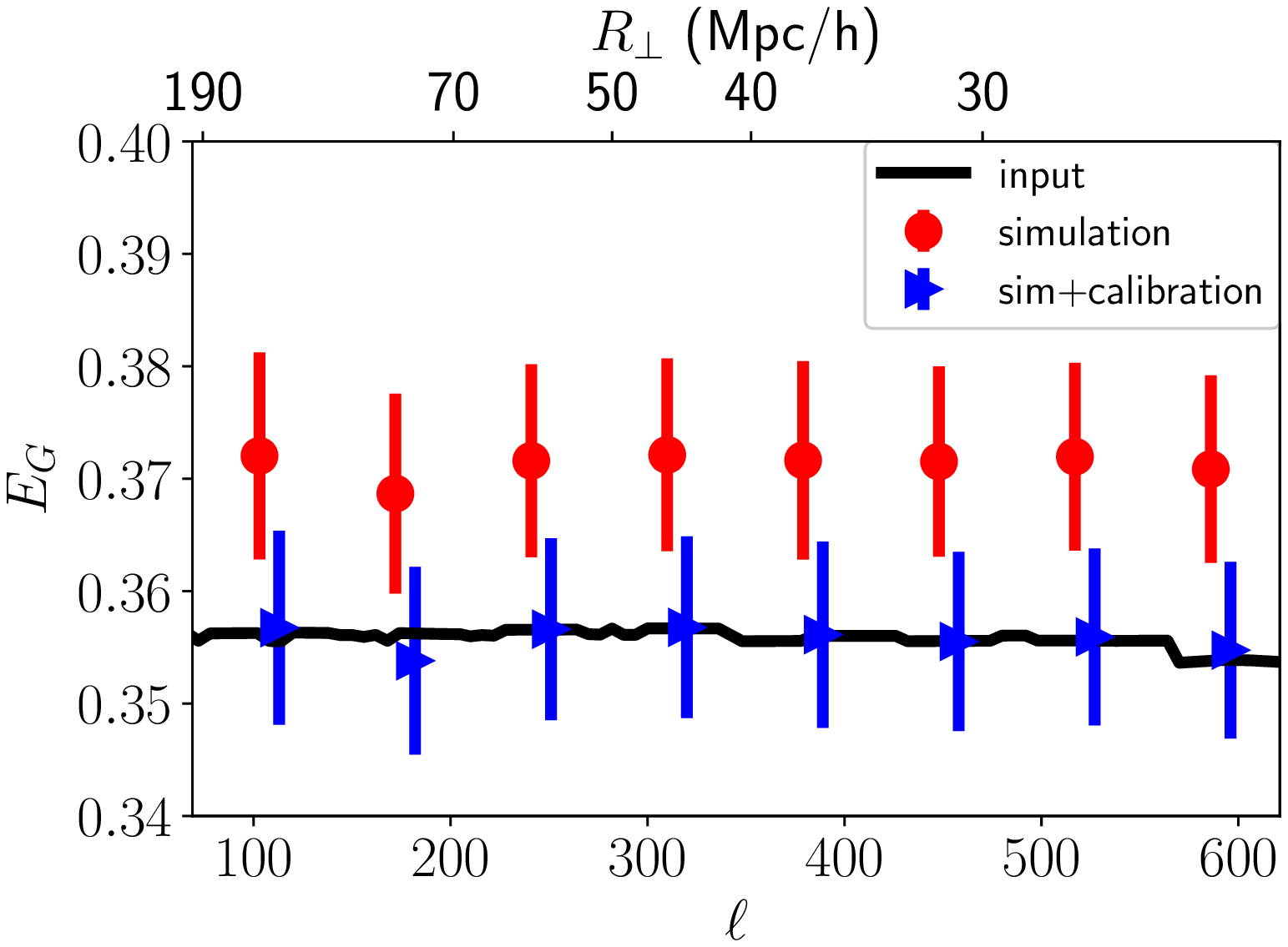}}\\
\end{tabular}		
\end{center}
\caption{Same as Figure \ref{fig:3} except for the photometric surveys. The rows from top to bottom corresponds to DES, LSST and Euclid (photo) surveys respectively. We shift data after calibration by $\Delta l=10$ for reading convenience.}
\label{fig:4b}
\end{figure*}

We now present results for the performance of our calibration method assuming that the parameters we use to construct the calibrations are known perfectly with no errors. We use 100 simulations in this analysis, and our results are shown in Figures \ref{fig:3} and \ref{fig:4b} and Tables \ref{t:2} and \ref{t:3}. Each error bar is the standard deviation of 100 simulations in one multipole bin in order to model the error for one realization. We find that for all spectroscopic and photometric survey cases the $E_G$ fractional deviation is less than $5\%$, while photometric survey cases suffer from a larger deviation on average. Ignoring shot noise from galaxy surveys and lensing noise from CMB surveys, the $E_G$ deviation in all spectroscopic surveys except DESI LRG and DESI ELG surveys cross-correlated with Adv.~ACT is much less than the error estimated by simulation.  In particular, DESI LRG and ELG may exhibit a magnification bias in $E_G$ that is greater than twice the statistical error.  This appears to be caused by DESI's high galaxy density and low clustering bias while maintaining a magnification bias parameter $s\neq0.4$.  For all the photometric surveys, the very high number density causes the magnification bias in $E_G$ to be approximately 4\%, much greater than statistical errors, even when cross-correlated with the Planck CMB lensing map. For these cases $E_G$ after calibration is corrected from about 2-4 times the simulation error to well within the errors. We believe this calibration will work increasingly well in future surveys with lower shot noise and lensing noise.\par
%In all galaxy surveys cross correlated with the Planck CMB survey, the lensing noise $N_l^{\kappa}$ is larger than the convergence auto power spectrum $C_l^{\kappa\kappa}$ and becomes the dominant noise source. For BOSS+eBOSS QSO and DESI QSO survey, since the redshift distribution is centered at a much higher redshift, the angular power spectrum amplitude is smaller and the shot noise from galaxy survey also becomes considerable. Thus, $E_G$ bias becomes unimportant in QSO and CMB cross correlation after including noise in the simulations. However, due to the large galaxy number density in DESI/Euclid surveys and low lensing noise in ACTPol, CMB-S4 survey, the $E_G$ error estimated by simulations with noise is still very small among cross correlation between these surveys, causing the $E_G$ deviation to be comparable to the simulation error.

%In the above figures, the first column shows $\Delta E_G/E_{g0}$, the second column shows simulations with CMB lensing noise from the Planck survey. The third column shows simulations with CMB lensing noise from Adv. ACTPol. The last column shows simulations with CMB lensing noise from CMB-S4. The rows from top to bottom corresponds to BOSS+eBOSS LRG, BOSS+eBOSS QSO, eBOSS ELG, DESI LRG, DESI ELG, DESI QSO and Euclid surveys respectively. We shift data after calibration by $\Delta l=10$ for reading convenience.

%In the above figures, the first column shows $\Delta E_G/E_{g0}$, the second column shows simulations with CMB lensing noise from the Planck survey. The third column shows simulations with CMB lensing noise from Adv. ACTPol. The last column shows simulations with CMB lensing noise from CMB-S4. The rows from top to bottom corresponds to DES, LSST and Euclid (photo) surveys respectively. We shift data after calibration by $\Delta l=10$ for reading convenience.

\begin{table*}
    \centering
    \begin{tabular}{c c c c c}
    \hline\hline
         &Theoretical $E_G$&no noise& Planck &Adv.~ACT  \\\hline
         %&$0.395$&$-0.68\%,2.30\%$&  $-1.22\%, 6.76\%$&$-0.88\%,2.79\%$&$-0.82\%,2.43\%$\\[-1ex]
         %\raisebox{1.5ex}{BOSS+eBOSS LRG}&$(0.92\%)$&$0.15\%,2.30\%$&$-0.40\%,6.71\%$&$-0.053\%,2.77\%$&$0.009\%,2.42\%$\\[1ex]\hline
         BOSS+eBOSS LRG&$0.390,0.89\%$&$-1.21\%,2.28\%$&  $-0.64\%, 7.13\%$&$-0.38\%,2.77\%$\\[1ex]\hline
         %&$0.332$&  $-0.91\%, 1.08\%$&\Ant{$4.29\%,1.26\%$}&$3.09\%,8.30\%$&$1.62\%,7.50\%$\\[-1ex]
         %\raisebox{1.5ex}{BOSS+eBOSS QSO}&$(1.39\%)$&$-0.056\%,1.07\%$&$3.27\%,10.43\%$&$3.26\%,7.12\%$&$1.91\%,6.54\%$\\[1ex]\hline
         BOSS+eBOSS QSO&$0.325,1.22\%$&  $-0.69\%, 1.08\%$&$2.52\%,13.71\%$&$2.98\%,8.09\%$\\[1ex]\hline
         %&$0.375$&  $1.01\%,4.07\%$&$-2.66\%, 18.78\%$&$1.22\%,.66\%$&$1.86\%,6.27\%$\\[-1ex]
         %\raisebox{1.5ex}{eBOSS ELG}&$(-1.15\%)$&$-0.045\%,4.07\%$&$-3.23\%,18.88\%$&$0.15\%,6.68\%$&$0.80\%,6.28\%$\\[1ex]\hline
         eBOSS ELG&$0.369,-1.12\%$&  $0.80\%,4.13\%$&$4.02\%, 19.35\%$&$2.05\%,7.14\%$\\[1ex]\hline
         %&$0.378$&  $-0.62\%,1.28\%$&$-0.033\%, 3.82\%$&$-0.77\%,1.70\%$&$-0.60\%,1.65\%$\\[-1ex]
         %\raisebox{1.5ex}{DESI LRG}&$(0.48\%)$&$-0.18\%,1.28\%$&$0.41\%,3.81\%$&$-0.33\%,1.70\%$&$-0.16\%,1.64\%$\\[1ex]\hline
         
         &&&  $3.28\%, 3.92\%$&$2.71\%,1.64\%$\\[-1ex]
         \raisebox{1.5ex}{DESI LRG}&\raisebox{1.5ex}{$0.372,-3.26\%$}&\raisebox{1.5ex}{$2.95\%,1.21\%$}&$0.64\%,4.01\%$&$0.0032\%,1.67\%$\\[1ex]\hline
         %&$0.360$&$-2.31\%,0.70\%$&  $-2.13\%, 3.17\%$&$-2.10\%,1.09\%$&$-2.32\%,1.01\%$\\[-1ex]
         %\raisebox{1.5ex}{DESI ELG}&$(2.47\%)$&$-0.12\%,0.68\%$&$-0.055\%,3.03\%$&$0.075\%,1.05\%$&$-0.13\%,0.98\%$\\[1ex]\hline
         &&&  $-4.13\%, 3.21\%$&$-3.95\%,1.15\%$\\[-1ex]
         \raisebox{1.5ex}{DESI ELG}&\raisebox{1.5ex}{$0.353,4.43\%$}&\raisebox{1.5ex}{$-4.02\%,0.68\%$}&$-0.12\%,3.06\%$&$0.10\%,1.08\%$\\[1ex]\hline
         %&$0.345$&$-2.02\%,0.53\%$&  $-1.40\%, 7.66\%$&$-1.52\%,3.73\%$&$-1.21\%,3.54\%$\\[-1ex]
         %\raisebox{1.5ex}{DESI QSO}&$(2.28\%)$&$-0.023\%,0.52\%$&$0.28\%,6.60\%$&$0.37\%,3.44\%$&$0.67\%,3.29\%$\\[1ex]\hline
         DESI QSO&$0.339,0.628\%$&$-0.39\%,0.52\%$&  $0.10\%, 7.88\%$&$0.035\%,3.76\%$\\[1ex]\hline
         %&$0.371$&$0.010\%,0.66\%$&  $-0.42\%, 2.79\%$&$-0.068\%,0.96\%$&$0.011\%,0.89\%$\\[-1ex]
         %\raisebox{1.5ex}{Euclid}&$(0.019\%)$&$0.11\%,0.68\%$&$-3.17\%,2.90\%$&$0.031\%,0.98\%$&$0.11\%,0.92\%$\\[1ex]\hline
         Euclid&$0.365,0.88\%$&$-0.17\%,0.68\%$&  $0.13\%, 3.05\%$&$-0.52\%,1.02\%$\\[1ex]\hline
    \end{tabular}
    \caption{Summary of simulation results for spectroscopic surveys. Theoretical $E_G$ column shows the average of $E_G$ from theory in all ranges on the left and averaged $\Delta E_G/E_G$ on the right. $E_G$ values using cross-correlations between galaxy surveys and CMB surveys are averaged over 8 bins in the ranges displayed in Figure \ref{fig:3}.  The remaining columns display $(\hat{E}_g-E_G^{theory})/E_G^{theory},\sigma(\hat{E}_g)/E_G^{theory}$ for each case. The upper line in each row corresponds to the averaged simulation result before calibration, the lower line presents averaged simulation result after calibration using Eqs.~\ref{eq:30}-\ref{eq:32}. Among all the simulation results, except for DESI LRG and DESI ELG columns, $E_G$ fractional deviation is much smaller than $1\sigma$ before calibration, in which case the advantage of calibration is not obvious, so we don't show calibrated results for the purpose of conciseness. The ``no noise'' column corresponds to averaged results from simulations without considering galaxy shot noise and CMB instrumental noise.}
    \label{t:2}
\end{table*}

\begin{table*}
    \centering
    \begin{tabular}{c c c c c}
    \hline\hline
         &Theoretical $E_G$&no noise& Planck &Adv.~ACT  \\\hline
         %&&$4.01\%,1.76\%$&  $4.12\%, 3.70\%$&$3.84\%,1.86\%$&$4.14\%,1.75\%$\\[-1ex]
         %\raisebox{1.5ex}{DES}&\raisebox{1.5ex}{$0.378,-4.15\%$}&$-0.033\%,1.65\%$&$-0.020\%,3.35\%$&$-0.18\%,1.74\%$&$0.080\%,1.65\%$\\[1ex]\hline
         &&&  $3.97\%, 4.05\%$&$3.86\%,2.34\%$\\[-1ex]
         \raisebox{1.5ex}{DES}&\raisebox{1.5ex}{$0.372,-4.09\%$}&\raisebox{1.5ex}{$4.14\%,2.32\%$}&$-0.086\%,3.69\%$&$-0.12\%,2.22\%$\\[1ex]\hline
         %&&$4.26\%,0.75\%$&  $4.27\%, 1.62\%$&$4.28\%,0.79\%$&$4.28\%,0.76\%$\\[-1ex]
         %\raisebox{1.5ex}{LSST}&\raisebox{1.5ex}{$0.369,-4.32\%$}&$0.0077\%,0.71\%$&$0.021\%,1.47\%$&$0.021\%,0.74\%$&$0.018\%,0.72\%$\\[1ex]\hline
         &&&  $4.42\%, 1.83\%$&$4.20\%,1.05\%$\\[-1ex]
         \raisebox{1.5ex}{LSST}&\raisebox{1.5ex}{$0.362,-4.26\%$}&\raisebox{1.5ex}{$4.22\%,1.02\%$}&$0.18\%,1.66\%$&$0.023\%,0.99\%$\\[1ex]\hline
         %&&$4.32\%,0.65\%$&  $4.20\%, 1.64\%$&$4.47\%,0.69\%$&$4.37\%,0.65\%$\\[-1ex]
         %\raisebox{1.5ex}{Euclid (photo)}&\raisebox{1.5ex}{$0.362,-4.47\%$}&$-0.074\%,0.61\%$&$-0.17\%,1.47\%$&$0.060\%,0.64\%$&$-0.031\%,0.61\%$\\[1ex]\hline 
         &&&  $4.14\%, 1.68\%$&$4.40\%,0.86\%$\\[-1ex]
         \raisebox{1.5ex}{Euclid (photo)}&\raisebox{1.5ex}{$0.356,-4.40\%$}&\raisebox{1.5ex}{$4.33\%,0.84\%$}&$-0.18\%,1.52\%$&$0.062\%,0.81\%$\\[1ex]\hline
    \end{tabular}
    \caption{Same as Table \ref{t:2} except for photometric surveys. We see that these surveys generally have magnification biases over 4\% while the calibration removes it efficiently.}
    \label{t:3}
\end{table*}

\subsection{Calibration tests under parameter uncertainties}\label{S:test2}

Our calibration method depends on the parameters that construct the calibration matching the true values in nature perfectly.  In reality, statistical and systematic errors will prevent this; thus, it is important to test how well our calibration method works when any parameters deviate from the true values under reasonable assumptions.  The parameters that we consider for this test are cosmological parameters, GR modifications, redshift distribution parameters, and clustering and magnification biases.  Although several parameters can deviate at once, biasing our calibration in unforeseen ways, we will only consider cases where one parameter deviates at a time to get a sense of the effect of each deviation.\par
There are multiple ways to test if our calibration method is stable with respect to these uncertainties.  Intuitively, one would assume one set of theoretical parameters to calibrate $C_l$ and multiple sets of possible true parameters to do simulations. However, we will perform the equivalent yet less time-consuming method of generating simulations of $\hat{C}_l$ from one set of true parameters for our universe, then assume multiple sets of mis-identified theoretical parameters to construct calibrations $C_l$.  We consider various parameter mis-identification cases by setting a baseline with all parameters as the true values then shifting them to the upper and lower bound of its uncertainty in order to produce a biased calibration for $C_l$. In this section we use cosmological parameters with uncertainties $\Omega_m=0.309\pm 0.007$, $\Omega_bh^2=0.02226\pm0.00016$, $\Omega_ch^2=0.1193\pm0.0014$, $h=0.679\pm0.007$, $\Omega_k=0.0000\pm0.0020$ according to \cite{Alam2016}.  The upper and lower limit of each parameter is for the Planck + BOSS BAO survey. Assuming standard deviations for each parameter is proportional to $1/\sqrt{V}$, where $V$ is the survey volume, we apply $\sigma_{DESI ELG}=0.22\sigma_{BAO}$, $\sigma_{LSST}=0.11\sigma_{BAO}$, where $\sigma_{BAO}$ represents parameter uncertainties for Planck + BOSS BAO survey.  We assume general relativity for a baseline along with the redshift distributions from DESI ELGs and LSST we use in section \ref{S:calitest} and expectation values of these cosmological parameters are the true parameters and use them to produce $\hat{C}_\ell$ in simulation. 

We choose $f(R)$ gravity and Chameleon gravity to test how modified gravity can influence the calibration performance.  It was shown in \citet{Pullen2015} that for functions $\mu(k,z)$ and $\gamma(k,z)$ that modify the Poisson equation and introduce large-scale anisotropic stress, respectively, $E_G$ is modified according to
\begin{eqnarray}
    E_G=\frac{\Omega_{m0}\mu(k,a)[1+\gamma(k,a)]}{2f}\, ,
\end{eqnarray}
We set general relativity (GR) as the base parameters, \emph{i.e.} $\gamma=\mu=1$. For $f(R)$ gravity \citep{Carroll2004,Song2007,Tsujikawa2007,Hojjati2011}), the relevant values for the parameters are
\begin{equation}
\begin{split}
&\mu^{fR}(k,a)=\dfrac{1}{1-B^{f(R)}_0a^{s-1}/6}\cdot\left[\dfrac{1+(2/3)B^{f(R)}_0a^s\bar{k}^2}{1+(1/2)B^{f(R)}_0a^s\bar{k}^2}\right]\\
&\gamma^{fR}(k,a)=\dfrac{1+(1/3)B^{f(R)}_0a^s\bar{k}^2}{1+(2/3)B^{f(R)}_0a^s\bar{k}^2}\\
&\bar{k}=\dfrac{c}{H_0}k
\end{split}
\end{equation}
In this work we use $B_0^{f(R)}=1.36\times10^{-5}$ \citep{2015PhRvD..91j3503B,2015PhRvD..91f3008X,Alam2015a}, $s=4$ (not the magnification bias).\par 
For Chameleon gravity \citep{Khoury2004,Bertschinger2008,Hojjati2011} we set $\mu$ and $\gamma$ to
\begin{equation}
\begin{split}
&\mu^{Ch}(k,a)=\dfrac{1+\beta_1\lambda_1^2k^2a^s}{1+\lambda_1^2k^2a^s}\\
&\gamma^{Ch}(k,a)=\dfrac{1+\beta_2\lambda_2^2k^2a^s}{1+\lambda_2^2k^2a^s}\\
&\lambda_2^2=\beta_1\lambda_1^2\\
&\beta_2=\dfrac{2}{\beta_1}-1
\end{split}
\end{equation}
In this work we use $\beta_1=1.2$, $\lambda_1=\sqrt{c^2B_0^{Ch}/(2H_0^2)}$, where $B_0^{Ch}=0.4$.

We consider $\Omega_k\neq0$ cases to test the stability of the calibration under $\Omega_k$ mis-identification.   When the universe is not flat, the coordinate distance can be computed generally as:
\begin{equation}\label{eq:34}
  r=\left\{
  \begin{array}{@{}ll@{}}
    \chi, & \text{if}\ \Omega_k=0 \\
    \dfrac{c}{H_0\sqrt{\Omega_k}}\sinh\left[\dfrac{\sqrt{\Omega_k} H_0\chi}{c}\right], & \text{if}\ \Omega_k>0 \\
    \dfrac{c}{H_0\sqrt{-\Omega_k}}\sin\left[\dfrac{\sqrt{-\Omega_k}H_0\chi}{c}\right], & \text{if}\ \Omega_k<0 \\
  \end{array}\right.\, .
\end{equation} 
\par 

To test how redshift distribution parameters affect $E_G$ measurements, we shift center of the redshift distribution $\mu$ and FWHM $\sigma$ by 1\%, 5\% and 10\% respectively to compute theoretical $C_l$ for mis-identified galaxy survey redshift distribution.  Specifically speaking, we define
\begin{eqnarray}
\mu=\int f_g(z)z dz\left[\int f_g(z)dz\right]^{-1}\, ,
\end{eqnarray}
and apply $f_g(z)\rightarrow f_g(z\pm r\mu)$ where $r=1\%$,$5\%$ and $10\%$. Likewise, we apply $f_g(z)\rightarrow f_g(\mu+(z-\mu)(1+r))$ to study impact of FWHM $\sigma$.\par
\begin{figure*}
	\centering
		\includegraphics[width=0.45\textwidth]{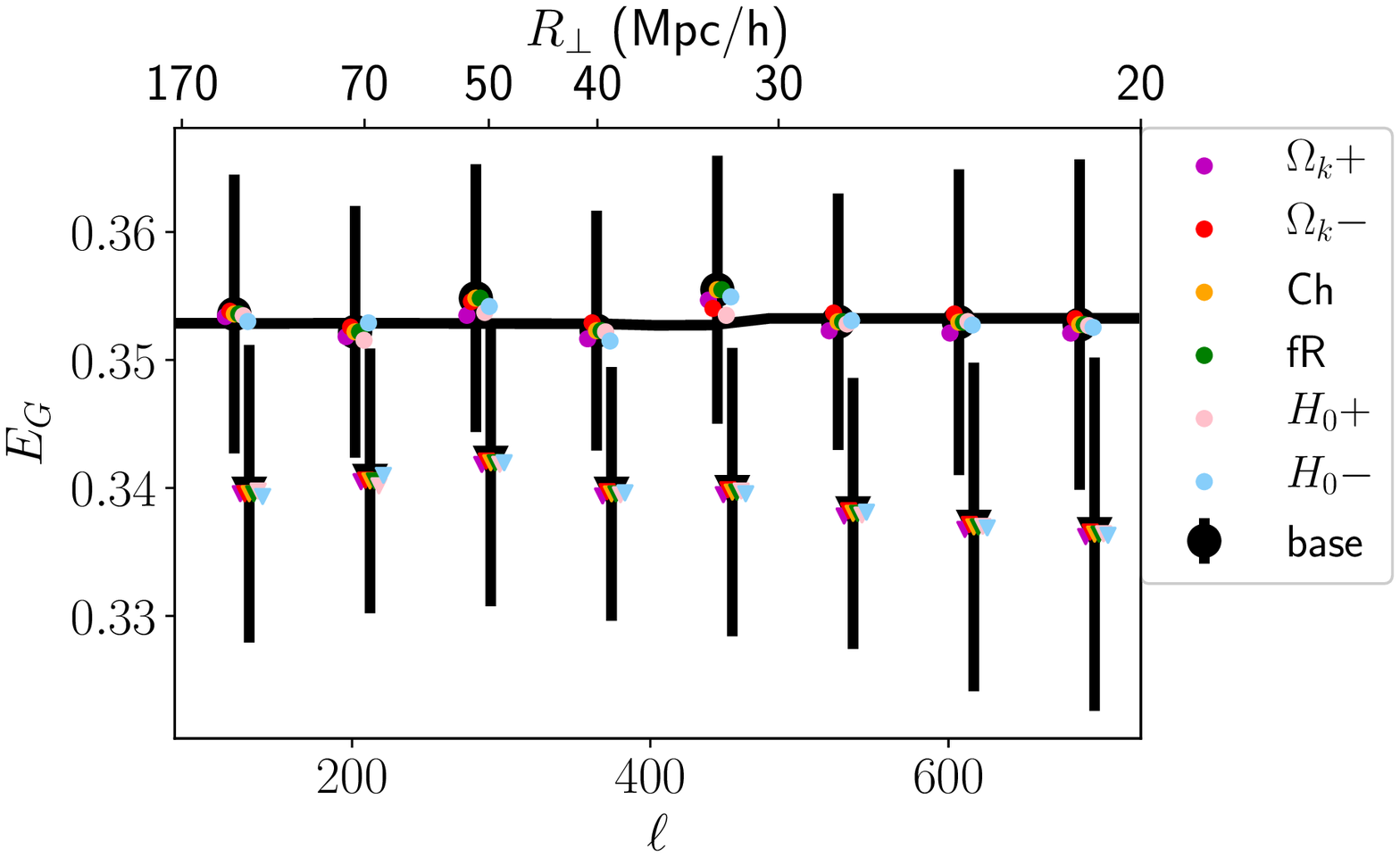}
		\includegraphics[width=0.45\textwidth]{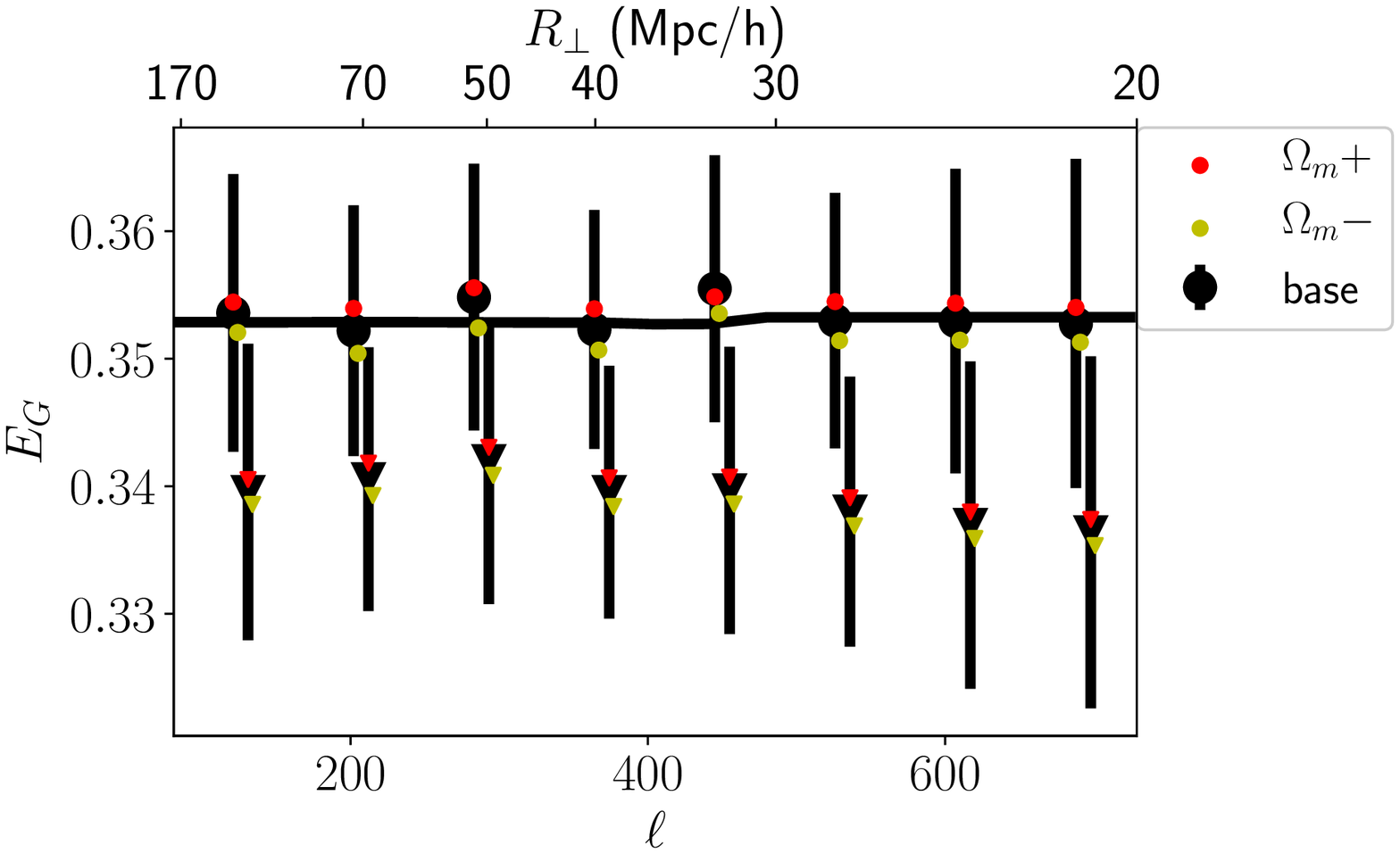}\\
		\includegraphics[width=0.45\textwidth]{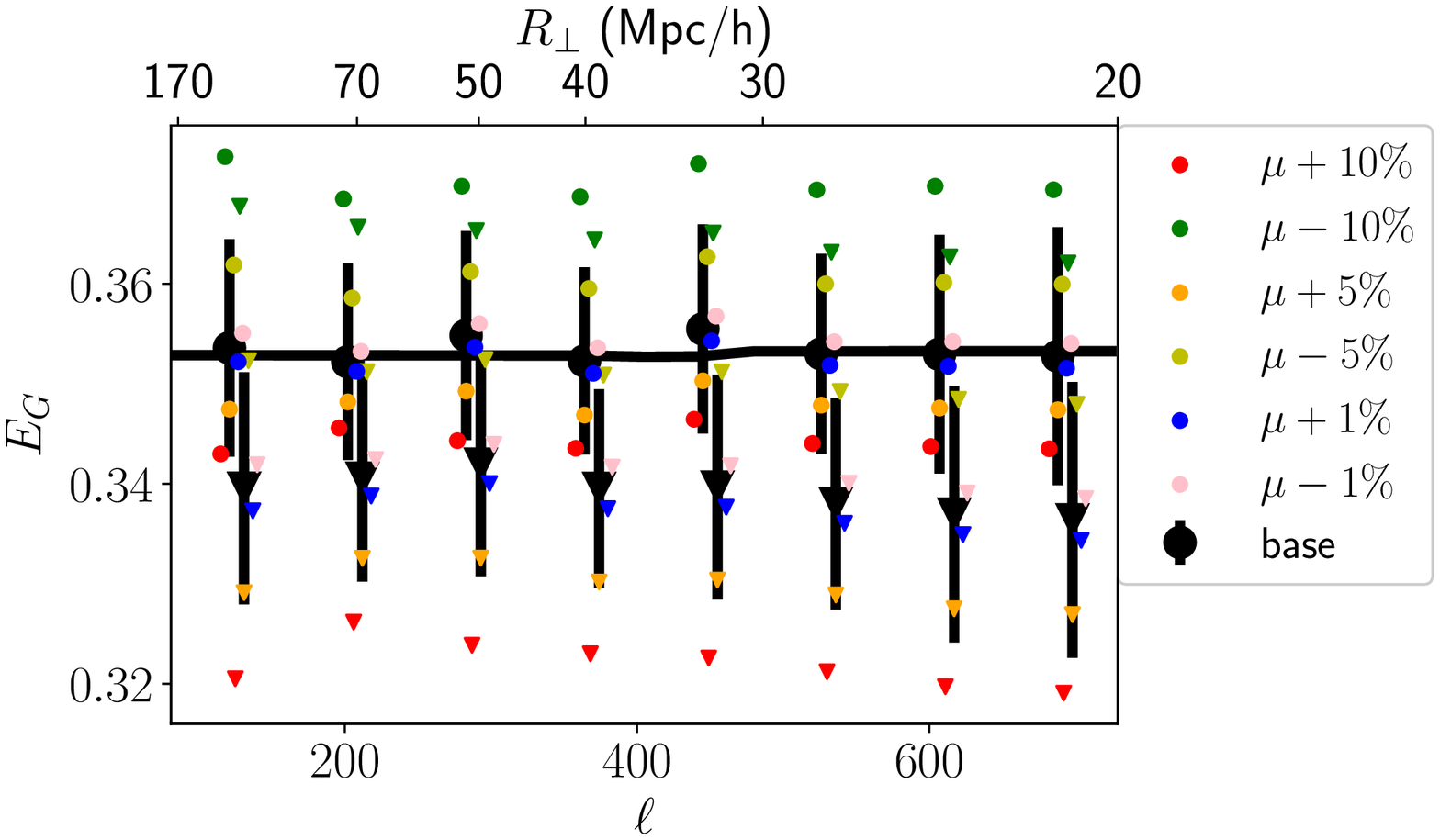}
		\includegraphics[width=0.45\textwidth]{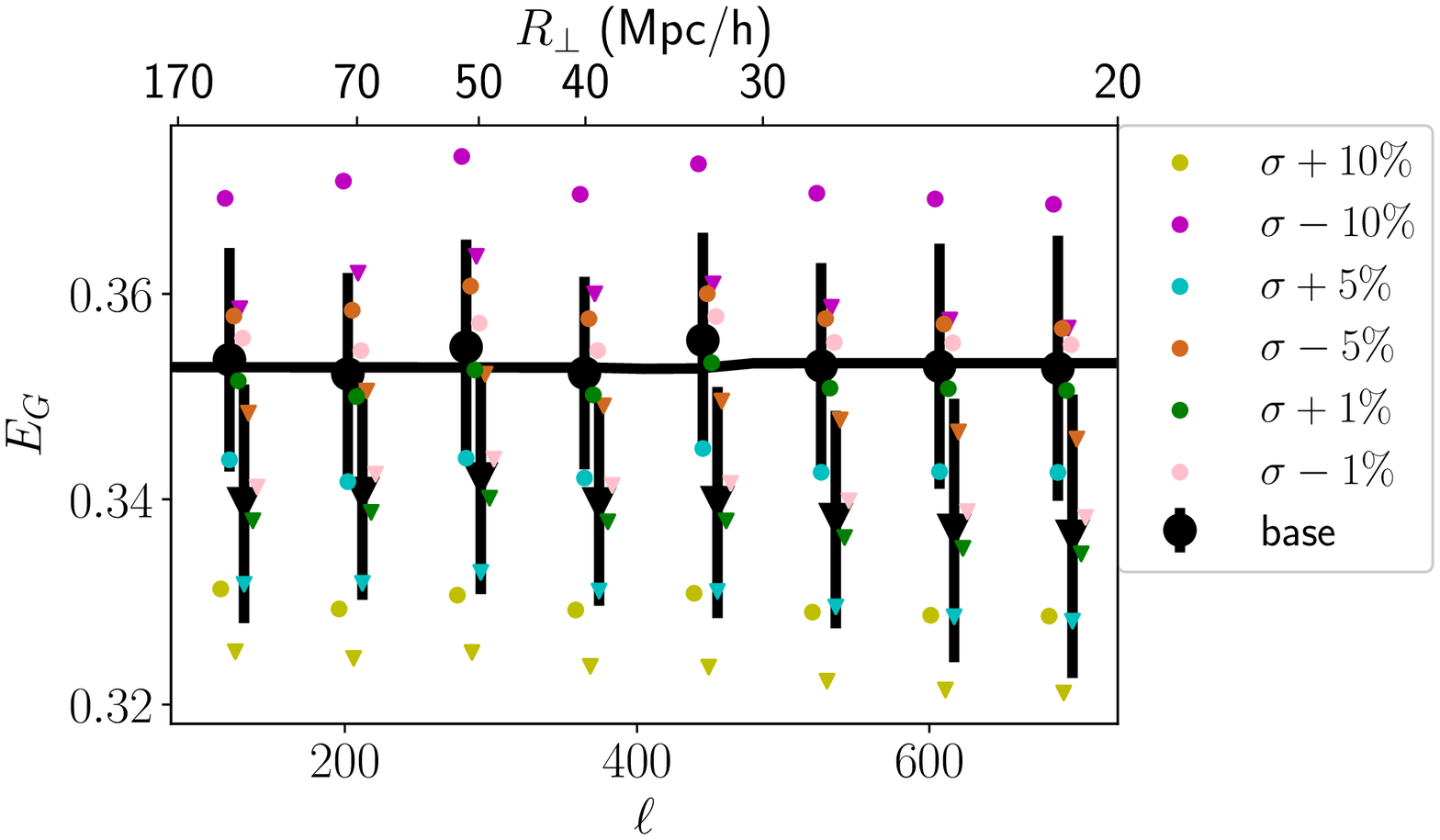}\\
		\includegraphics[width=0.45\textwidth]{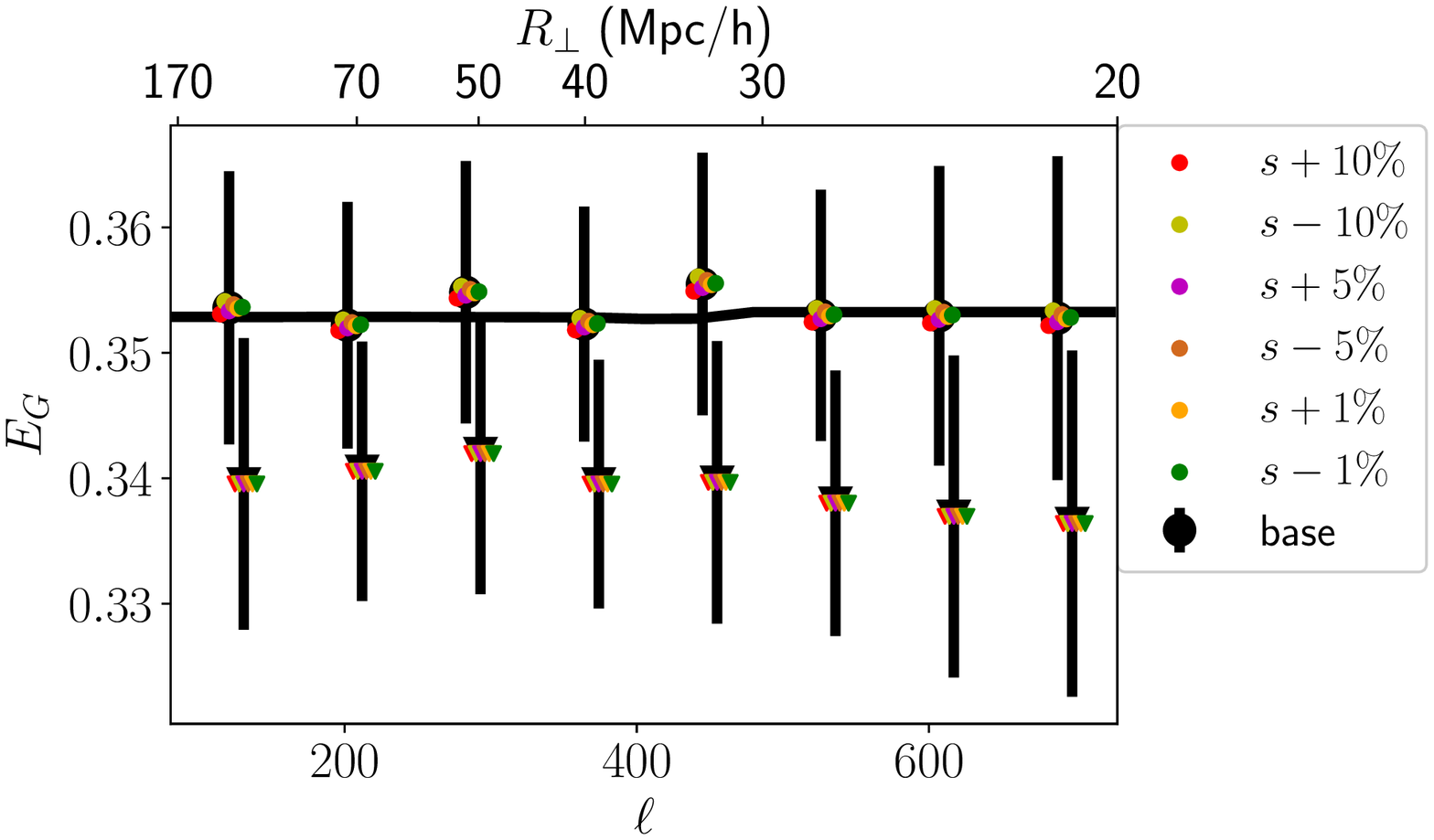}
		\includegraphics[width=0.45\textwidth]{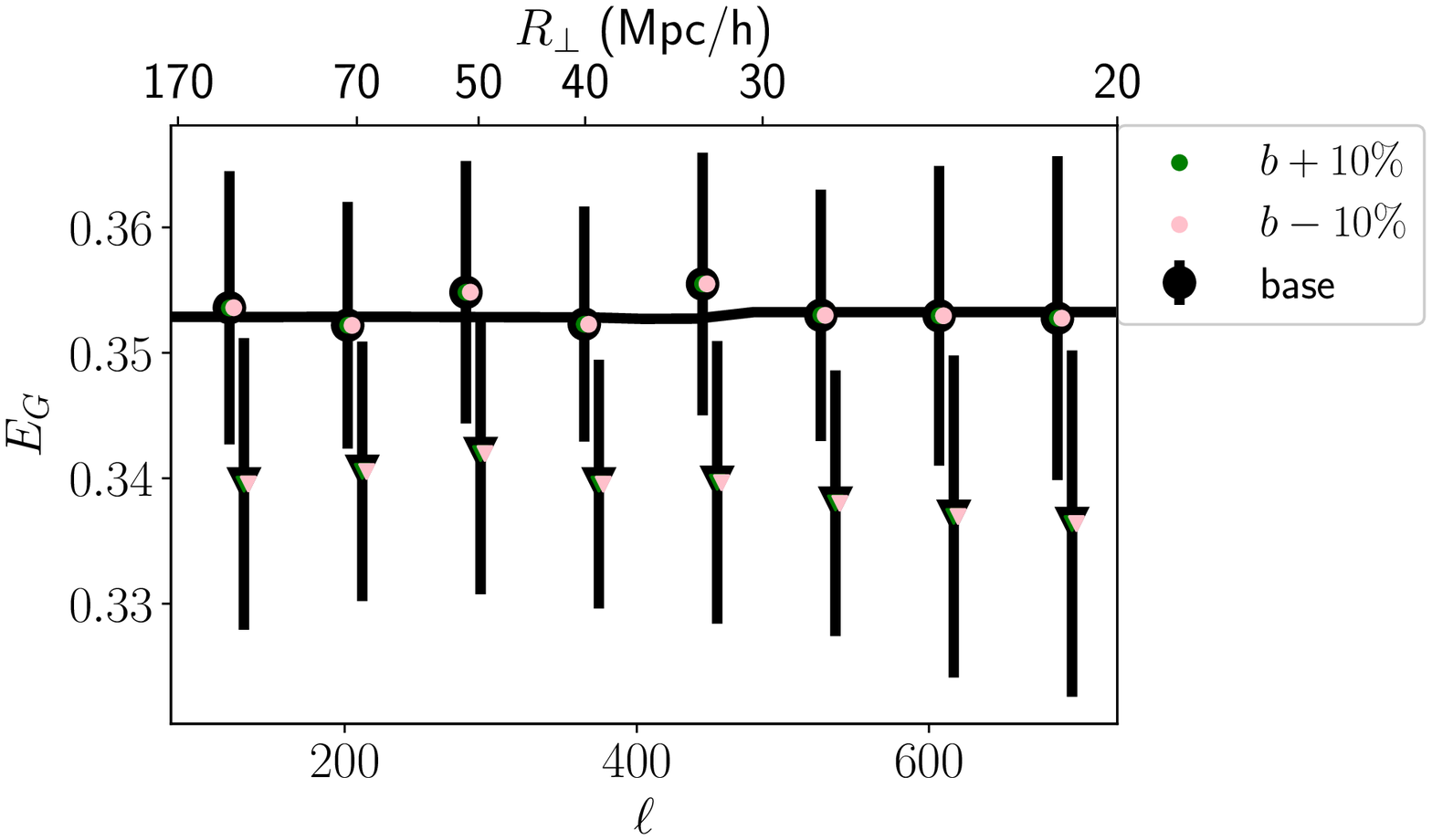}\\
	\caption{Calibration under cosmological parameters, redshift distribution parameters and GR modification mis-identification for DESI ELG cross Adv.~ACT. The dark line is the true $E_G$ value and dark data points are estimated from true parameters. Data marked by triangle are simulation results before calibration. Data marked with circle are simulation results after calibration. In the legend, '+/-' sign means we use upper/lower bound of parameter to compute theoretical angular power spectrum. We shift different data sets by certain amount in order to see them clearly.}
	\label{fig:4}
\end{figure*}

\begin{figure*}
	\centering
		\includegraphics[width=0.45\textwidth]{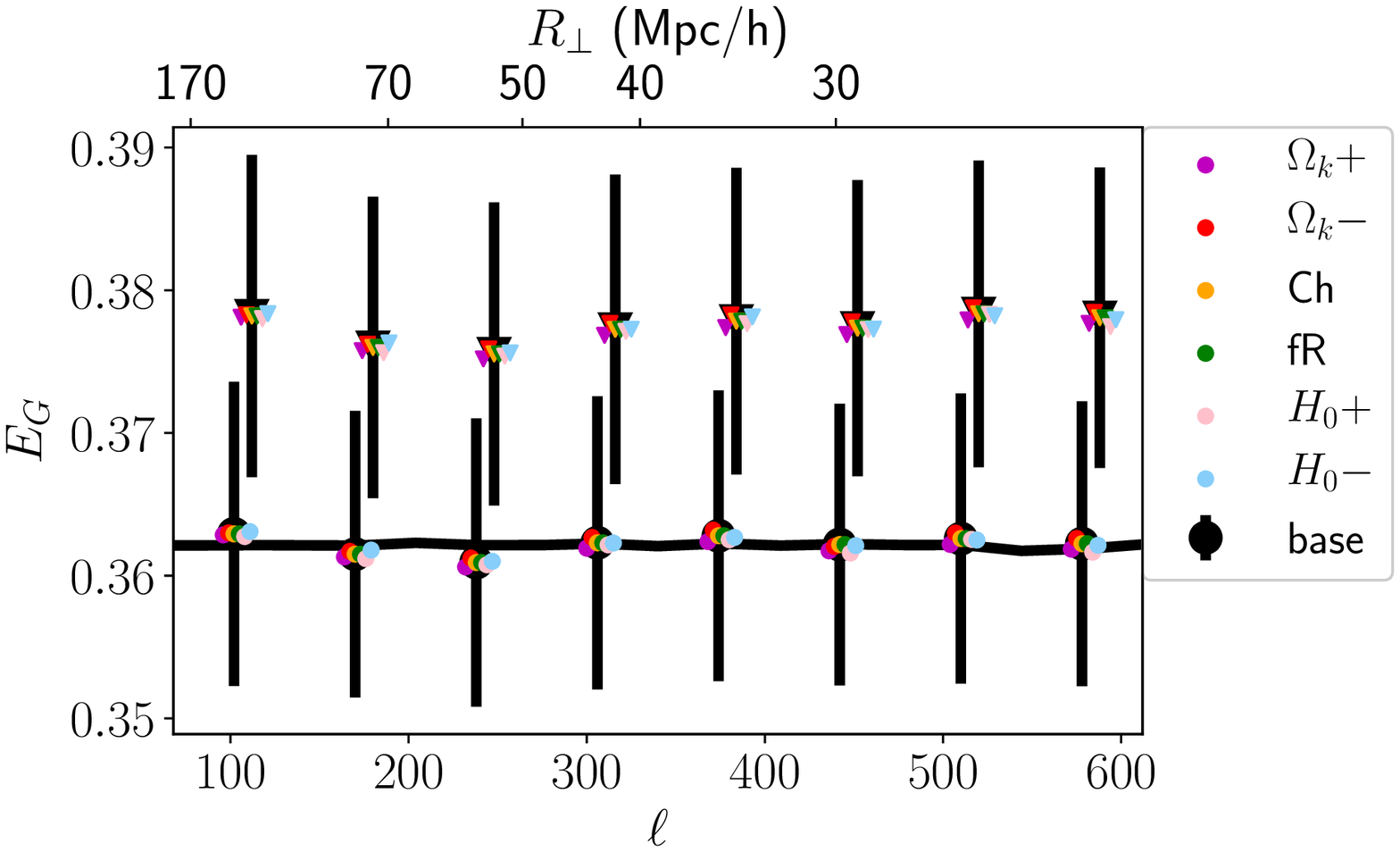}
		\includegraphics[width=0.45\textwidth]{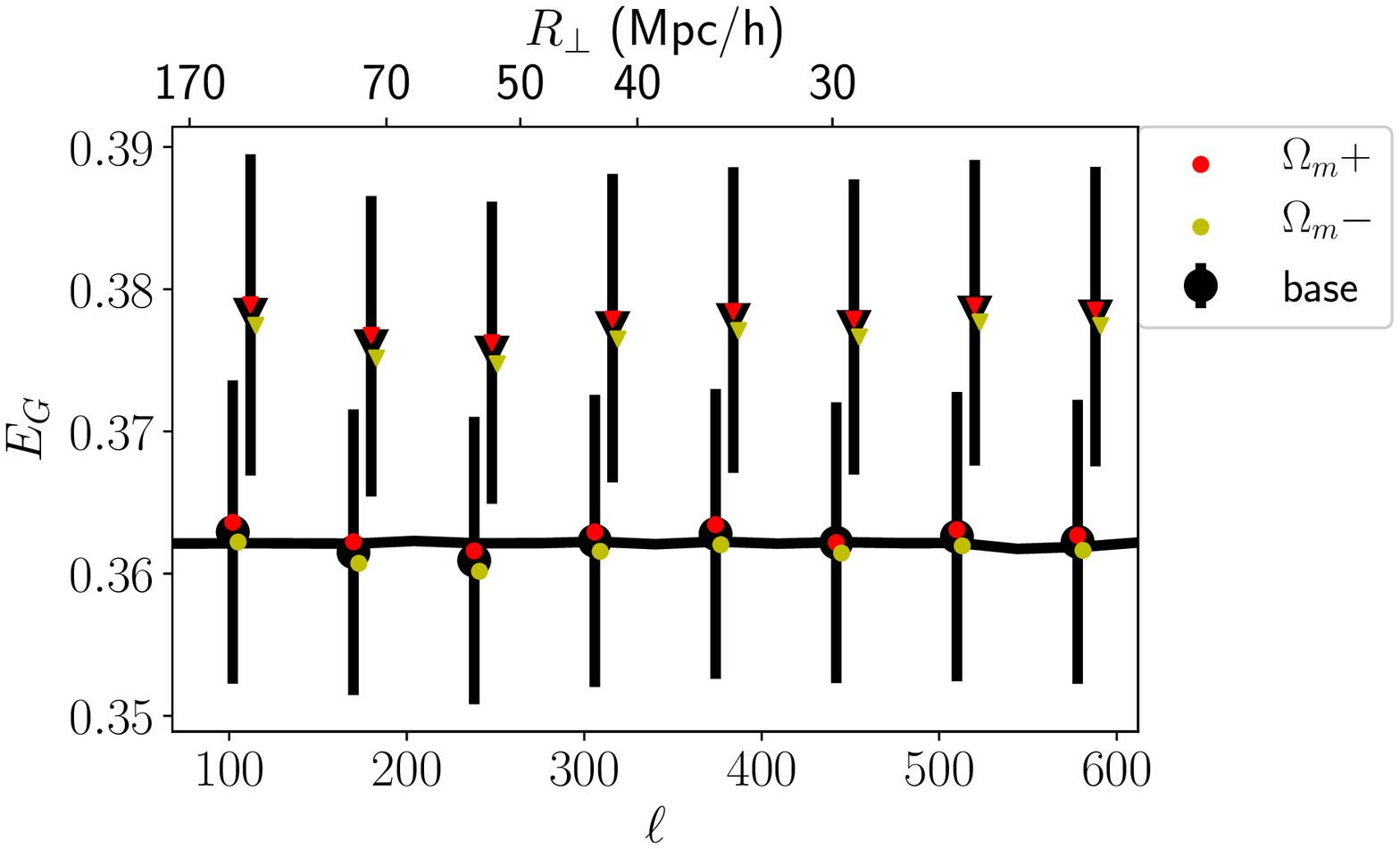}\\
		\includegraphics[width=0.45\textwidth]{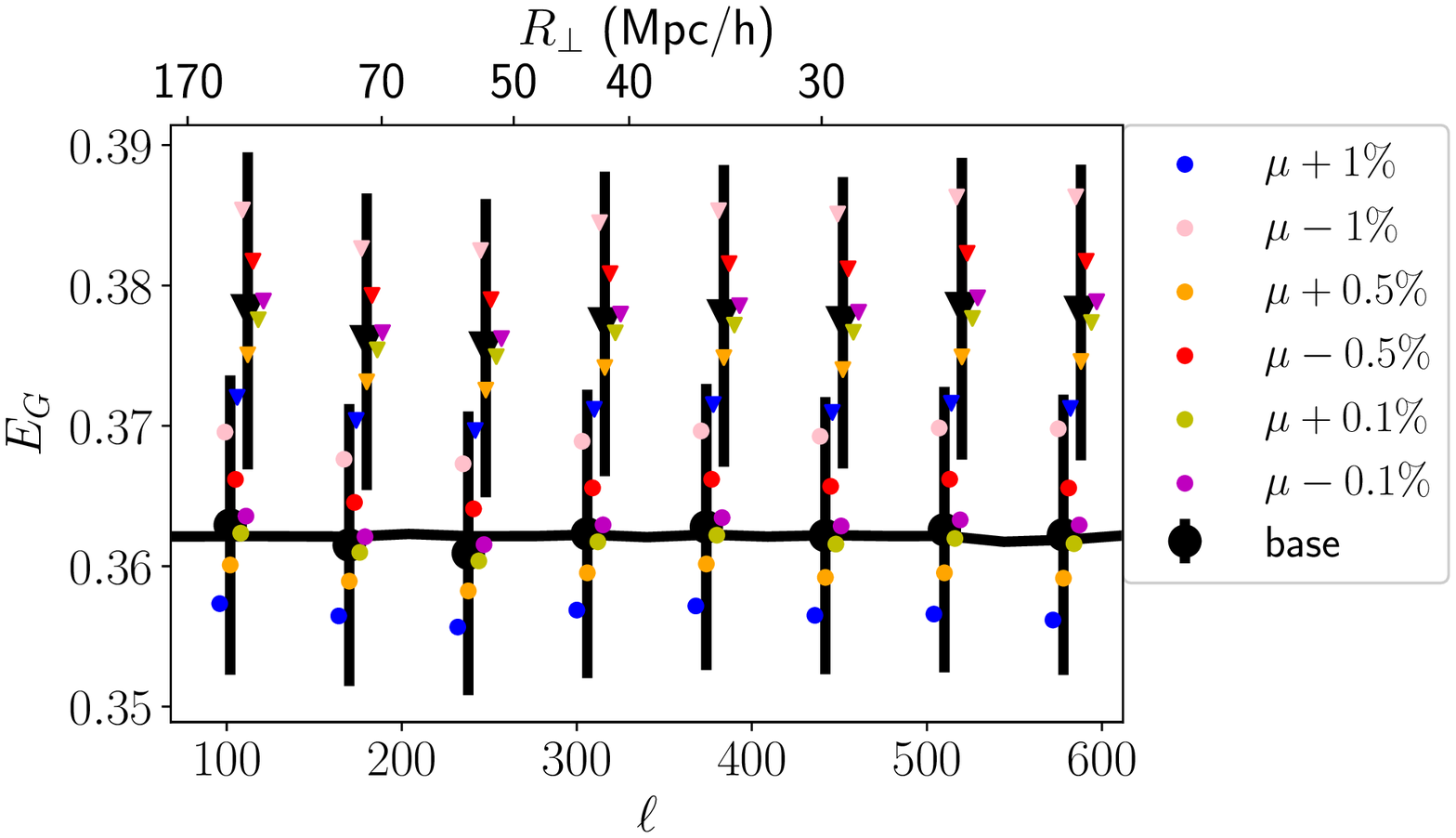}
		\includegraphics[width=0.45\textwidth]{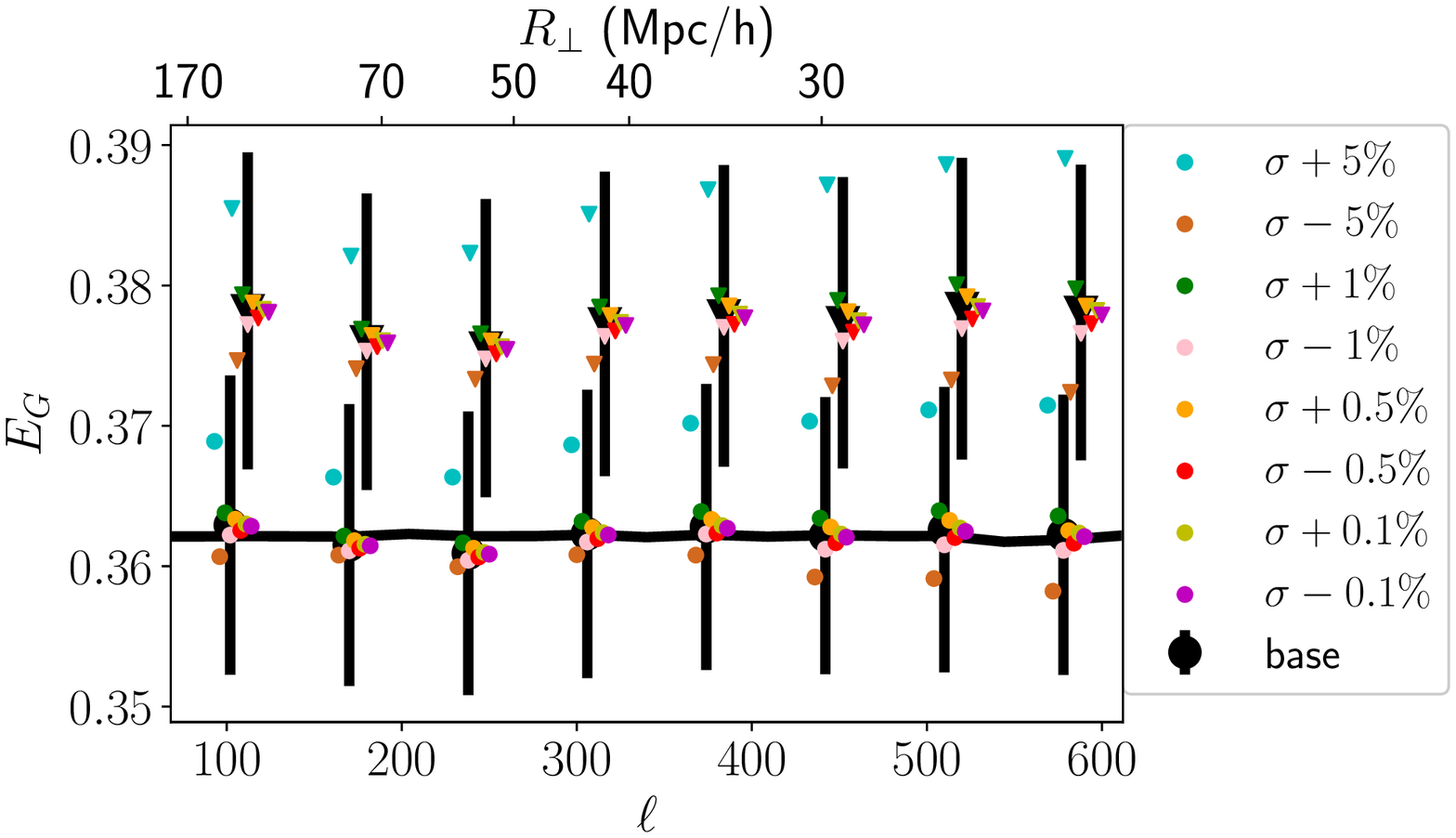}\\
		\includegraphics[width=0.45\textwidth]{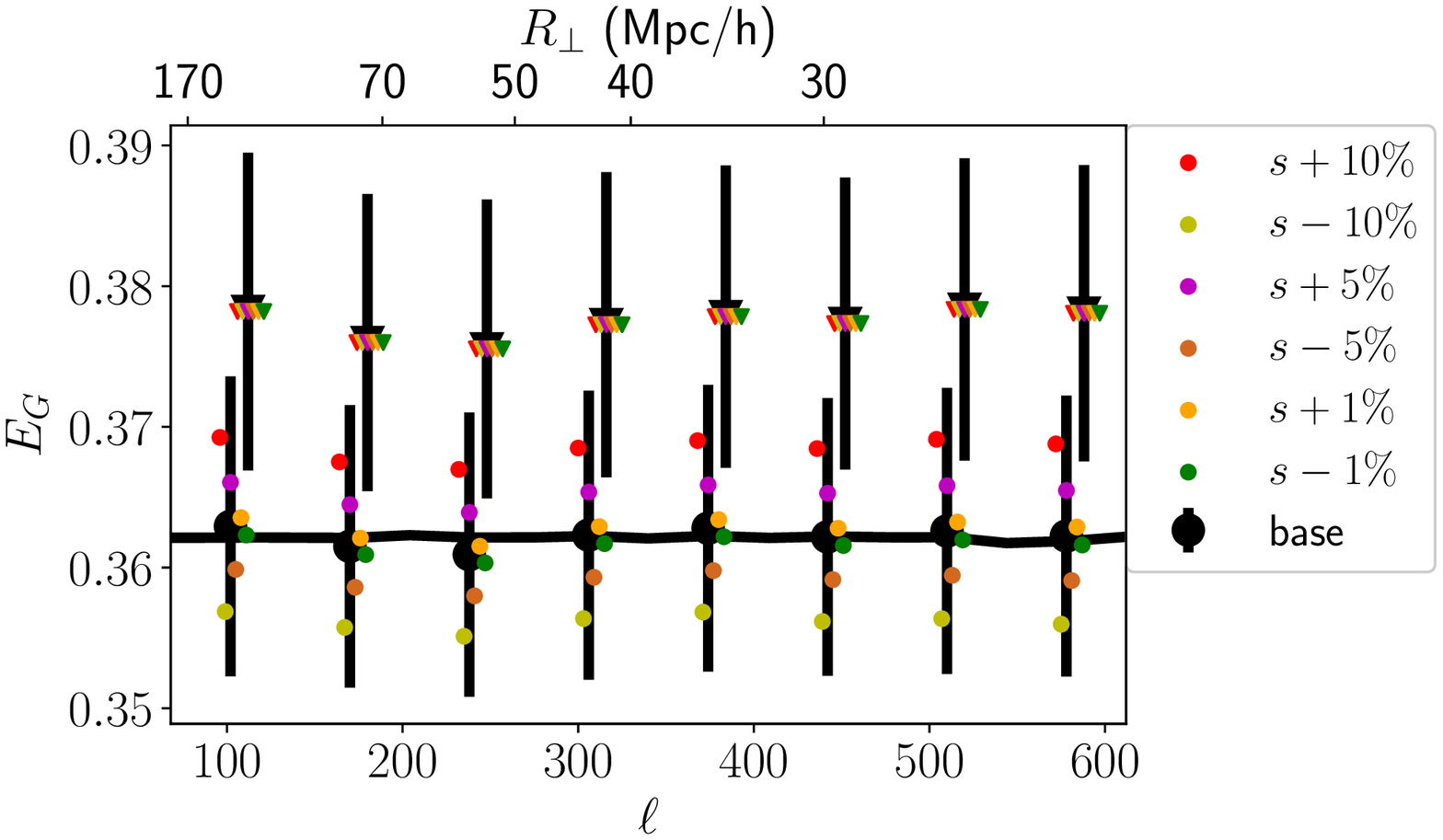}
		\includegraphics[width=0.45\textwidth]{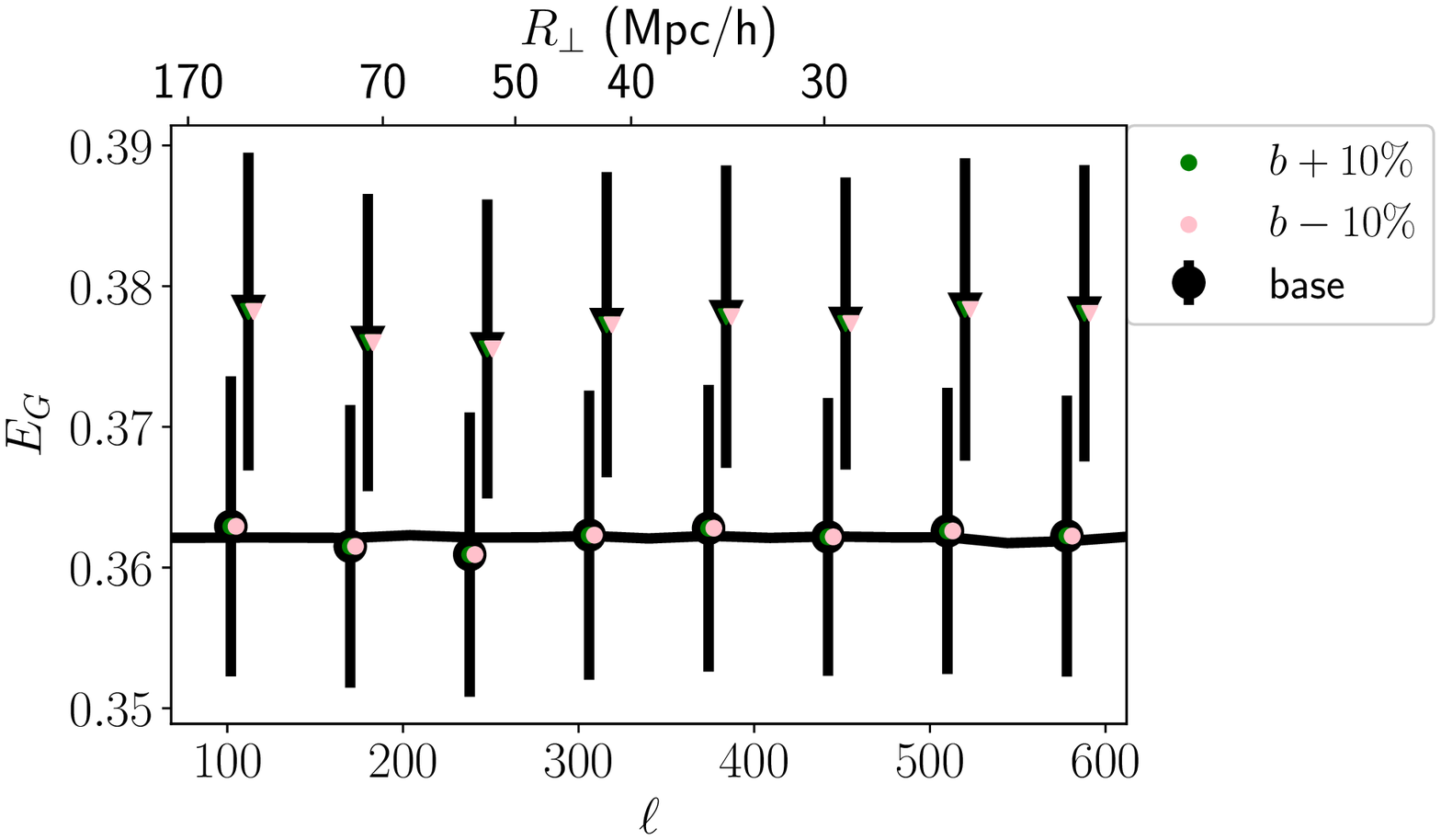}\\
	\caption{Calibration under cosmological parameters, redshift distribution parameters and GR modification mis-identification for LSST cross Adv.~ACT. Notation conventions are same with Figure \ref{fig:4}.}
	\label{fig:5}
\end{figure*}

 
In this discussion we will use shorthand 'A cross B' to refer to 'survey A cross-correlated with survey B'. Taking DESI ELG cross Adv.~ACT and LSST cross Adv.~ACT as examples, we apply our calibration for all parameter mis-identification cases and the results are shown in Figure \ref{fig:4} and \ref{fig:5}. Data marked by triangle is $E_G$ estimated from simulations where our calibration is not applied. Data marked by circles are calibrated estimation under mis-identified parameters. Data in black are results from true parameters.\par
For DESI, we find that our calibration method is stable with respect to shifts in most cosmological parameters, as well as for shifts in both the magnification and clustering biases.  However, the parameters for the redshift distribution will need to be known at the 1\%-level. LSST has a similar story, except that the magnification bias parameter $s$ will also need to be know with 1\% precision.  The redshift distribution and the magnification bias parameter can be measured directly from the galaxy sample, thus knowing these values precisely enough should be feasible. \par
Gravity determines redshift space distortion parameter $\beta$ and brings $\mu(k,a)(\gamma(k,a)+1)/2$ factors to angular power spectra $C_\ell$, so it has significant influence on $E_G$. However, since $\beta$ is a measurable quantity and  $\mu(k,a)(\gamma(k,a)+1)/2$ factors are very close to unity for $f(R)$ and Chameleon gravity, our calibration method is not sensitive to the tested GR modification.\par

%\Ant{Why? even under $s\pm10\%$ the calibrated result is closer to the true value.}
%under the current uncertainty of parameters $\Omega_k$, $H_0$ and GR modifications. This means $C^{\kappa g}_{l2}/C^{\kappa\kappa}_{l}$, $C^{gg}_{l22}/C^{\kappa\kappa}_l$ and $C^{gg}_{l12}/C^{\kappa g}_{l1}$ in Eq (\ref{eq:30}-\ref{eq:32}) do not change a lot under variation of $\Omega_k$, $H_0$ and GR modification. However, it may lose interest if matter density $\Omega_m$, redshift distribution location $\mu$ and FWHM $\sigma$ are mis-identified since uncertainty of these parameters, which we do not consider when estimating the error, brings larger deviation to $E_G$ compared to the bias caused by lensing magnification effect.
\par

\par

\section{Conclusions}\label{S:conclu}

The $E_G$ statistic is a promising tool to probe GR deviation. \cite{MoradinezhadDizgah2016a} questions that $E_G$ may be biased up to $40\%$ due to the lensing magnification effect in surveys collecting high redshift information and therefore loses its interest.\par 
In this work, we show that the lensing magnification is not as significant as \cite{MoradinezhadDizgah2016a} proposed. We also show that for BOSS+eBOSS and upcoming DESI, Euclid survey the $E_G$ fractional bias is less than $5\%$. However, this bias is not necessarily small compared to measurement error due to low shot noise and lensing noise in future surveys. We therefore propose a new calibration method -- using the convergence auto power spectrum $C_{\kappa\kappa}$ to estimate the higher order terms contributed by lensing magnification effect. We find the advantage of this calibration is not obvious for surveys with large noise and large growth rate uncertainty. This method calibrates $E_G$ significantly in DESI LRG and ELG cross correlate with Adv.~ACT CMB lensing since lensing noise in these surveys are much lower. Since photometric surveys contain greater galaxy number density compared to spectroscopic surveys, which largely suppresses the shot noise, this method can calibrates $E_G$ effectively even for photometric surveys cross correlate with Planck CMB lensing. Of course the calibration performance gets better for Adv. ~ACT CMB lensing cases. Moreover, this method works stably under variance of $H_0$, $\Omega_k$ and GR modification, but is sensitive to errors in the redshift distribution and the magnification bias parameter $s$ which are feasible to maintain at a minimal level. This calibration method will be a promising solution to $E_G$ bias due to lensing magnification effect as the precision of future surveys keep increasing.

\section*{Acknowledgments}
We wish to thank A.~Dizgah for discussion about the formalism, as well as M.~Blanton, S.~Ho, and J.~Tinker for comments on our manuscript.

\appendix

\section{Limber approximation expressions for $C_\ell$}\label{S:cllimber}
Here are the expressions we use for the Limber approximation for the various $C_\ell$'s:
\begin{eqnarray}\label{eq:24}
    C_\ell^{g1g1}=\int_0^{\infty}dz\frac{H(z)}{cr^2(z)}f_g^2(z)b^2(z)P\left[\frac{\ell}{r(z)},z\right]
\end{eqnarray}
\begin{eqnarray}\label{eq:25}
    C_\ell^{\kappa g1}&=&\frac{3H_0^2\Omega_{m0}}{2c^2}\int_0^{\infty}dz\frac{W(z_{CMB},z)}{r^2(z)}f_g(z)\nonumber\\
    &&\times(1+z)b(z)P\left[\frac{\ell}{r(z)},z\right]\nonumber\\
    &&\times \dfrac{\mu(k,z)(\gamma(k,z)+1)}{2}
\end{eqnarray}
\begin{eqnarray}\label{eq:26}
C_\ell^{g1g2}&=&\frac{3\Omega_{m0}H_0^2}{c^2}\left(\dfrac{5}{2}s-1\right)\int_0^{\infty}dz_1f_g(z_1)\int_0^{z_1}dz_2\nonumber\\
&&\times\frac{W(z_1,z_2)}{r^2(z_2)}(1+z_2)f(z_2)b(z_2)\nonumber\\
&&\times P\left[\frac{\ell}{r(z_2)},z_2\right]\dfrac{\mu(k,z_2)(\gamma(k,z_2)+1)}{2}
\end{eqnarray}
\begin{eqnarray}\label{eq:27}
C_\ell^{g2g2}&=&\left(\frac{3\Omega_{m0}H_0^2}{c^2}\right)^2\left(\dfrac{5}{2}s-1\right)^2\int_0^{\infty}dz_1f_g(z_1)\nonumber\\
&&\times \int_0^{\infty}dz_2f_g(z_2)\int_0^{min[z_1,z_2]}\frac{cdz_3}{H(z_3)}(1+z_3)^2\nonumber\\
&&\times \frac{W(z_1,z_3)W(z_2,z_3)}{r^2(z_3)}P\left[\frac{\ell}{r(z_3)},z_3\right]\nonumber\\
&&\times\left(\dfrac{\mu(k,z_3)(\gamma(k,z_3)+1)}{2}\right)^2
\end{eqnarray}
\begin{eqnarray}\label{eq:28}
C_\ell^{\kappa g2}&=&\frac{1}{2}\left(\frac{3\Omega_{m0}H_0^2}{c^2}\right)^2\left(\dfrac{5}{2}s-1\right)\int_0^{\infty}dz_1f_g(z_1)\nonumber\\
&&\times\int_0^{min[z_1,z_{CMB}]}dz_2\frac{c}{H(z_2)}(1+z_2)^2\nonumber\nonumber\\
&&\times \frac{W(z_1,z_2)W(z_{CMB},z_2)}{r^2(z_2)}P\left[\frac{\ell}{r(z_2)},z_2\right]\nonumber\\
&&\times\left(\dfrac{\mu(k,z_2)(\gamma(k,z_2)+1)}{2}\right)^2
\end{eqnarray}
\begin{eqnarray}\label{eq:29}
C_l^{\kappa\kappa}&=&\frac{1}{4}\left(\frac{3\Omega_{m0}H_0^2}{c^2}\right)^2\int_0^{z_{CMB}}dz\frac{c}{H(z)}\frac{W^2(z_s,z)}{r^2(z)}\nonumber\\
&&\times(1+z)^2P\left[\frac{\ell}{r(z)},z\right]\nonumber\\
&&\times\left(\dfrac{\mu(k,z)(\gamma(k,z)+1)}{2}\right)^2
\end{eqnarray}
here $r(z)$ is given by Eq. \ref{eq:34}, $P[\ell/r(z),z]=P[k=\ell/r(z),z]$ and $z_{CMB}=1100$ is the CMB redshift.

\section{magnification bias $s$ and Schechter Luminosity function }\label{S:mag_fac}
Following definition of \cite{2009A&A...507..683H}, lensed cumulative number counts of galaxies $N(>f)$ relates to the original number counts $N_0(>f)$ as:
\begin{equation}
    N(>f)=\mu^{-1}N_0(>\mu^{-1}f)
\end{equation}\par
Assume $N_0(>f)$ can be approximated by power law:
\begin{equation}\label{eq:b2}
    N_0(>f)=Af^{-\alpha}
\end{equation}
\begin{equation}\label{eq:b3}
\begin{split}
    N(>f)&=\mu^{-1}A\mu^{\alpha}f^{-\alpha}=\mu^{\alpha-1}N_0(>f)\\
    &\Longrightarrow \dfrac{dlogN(>f)}{dlogf}=-\alpha
\end{split}
\end{equation}
Notice the magnitude $m$ relates to flux $f$ as:
\begin{equation}\label{eq:b4}
\begin{split}
    m&=-2.5log(f)+Const.\\
    \Longrightarrow dm&=-2.5dlog(f)\\ \Longrightarrow \alpha&=-\dfrac{dlogN(>f)}{dlogf}=2.5\dfrac{dlogN(<m)}{dm}=2.5s
\end{split}
\end{equation}
Given relation between lensing factor $\mu$ and lensing convergence $\kappa$: $\mu\approx(1-\kappa)^{-2}$ \cite{Singh2018}, Eq. \ref{eq:b3} becomes:
\begin{equation}
\begin{split}
    N(>f)&\approx (1-\kappa)^{-2(\alpha-1)}N_0(>f)\\
    &\approx[1+2(\alpha-1)\kappa]N_0(>f)
\end{split}
\end{equation}\par
Now it is easy to express the lensed overdensity field to the first order:
\begin{equation}
\begin{split}
    \delta_g&=\dfrac{N-\Bar{N}}{\Bar{N}}\approx\dfrac{N_0-\Bar{N}+2(\alpha-1)\kappa N_0}{\Bar{N}}\\
    &\approx\delta_{g0}+2(\alpha-1)\kappa=\delta_{g0}+(5s-2)\kappa
\end{split}
\end{equation}\par
On the other hand, many works follow the definition of luminosity function on \cite{1976ApJ...203..297S}:
\begin{equation}\label{eq:b7}
    \Phi(L)dL=\Phi^*\left(\dfrac{L}{L^*}\right)^{\alpha_{LF}}exp\left(-\dfrac{L}{L^*}\right)d\left(-\dfrac{L}{L^*}\right)
\end{equation}
Here $\Phi(L)$ is the number density of galaxies with luminosity $L$. To compare Eq. \ref{eq:b7} with Eq. \ref{eq:b2} and find relation between $\alpha$ and $\alpha_{LF}$, let me transfer $\Phi(L)$ into cumulative number density $n(>f)$ through:
\begin{equation}
    n(>f)=\int^\infty_{4\pi r^2(\Bar{z})f}\Phi(L)dL=\Phi^*\Gamma(1+\alpha_{LF},\dfrac{4\pi fr^2(\Bar{z})}{L^*})
\end{equation}
\begin{equation}\label{eq:b9}
    lim_{f\rightarrow 0}n(>f)\rightarrow -\dfrac{\Phi^*}{1+\alpha_{LF}}\left(\dfrac{f\pi r^2(\Bar{z})}{L^*}\right)^{1+\alpha_{LF}}
\end{equation}
\par
Since we are only studying a shell located at redshift $\bar{z}$, cumulative galaxy number counts $N(>f)$ and number density $n(>f)$ are proportional to each other. Compare Eq. \ref{eq:b3} with Eq. \ref{eq:b9}, we have shown that $\alpha=-(1+\alpha_{LF})$.\par

\bibliographystyle{mnras}
%\bibliography{cut_off_lens_in_Eg,mybib}
\bibliography{Eg}
\label{lastpage}
\end{document}